\documentclass[11pt]{article}
\usepackage{amssymb,amsmath,amsfonts}
\usepackage{graphicx}
\usepackage{graphics}
\usepackage{eepic,epsfig}

\begin{document}

\title{Vacuum polarization induced by a cosmic string and a brane in AdS spacetime}
\author{W. Oliveira dos Santos$^{1}$%
\thanks{%
		E-mail: wagner.physics@gmail.com} ,\thinspace\ E. R. Bezerra de Mello$^{1}$\thanks{%
		E-mail: emello@fisica.ufpb.br} \\
	\\
	$^{1}$\textit{Departamento de F\'{\i}sica, Universidade Federal da Para\'{\i}%
		ba}\\
	\textit{58.059-970, Caixa Postal 5.008, Jo\~{a}o Pessoa, PB, Brazil}}
\maketitle

\begin{abstract}
In this paper we investigate the vacuum polarization effects associated to a charged quantum massive scalar field on a $(D+1)$-dimensional anti-de Sitter background induced by a magnetic-flux-carrying cosmic string in the braneworld model context. We consider the brane parallel to the anti-de Sitter boundary and the cosmic string orthogonal to them. Moreover, we assume that the field obeys the Robin boundary condition on the brane. Because the brane divides the space into two regions with different properties of the quantum vacuum, we calculate the vacuum expectation value (VEV) of the field squared and the energy-momentum tensor (EMT) in each region. To develop these analyses, we have constructed the positive frequency Wightman function for both regions. The latter is decomposed in a part associated with the anti-de Sitter bulk in the presence of a cosmic string only, and the other part induced by the brane. The vacuum polarization effects associated with the higher-dimensional anti-de Sitter bulk in the presence of cosmic string have been developed in the literature, and here we are mainly interested in the effects induced by the brane. We show that the VEVs of the field squared and the components of the EMT induced by the cosmic string are finite on the brane. Explicitly, we compare these observables with the corresponding ones induced by the brane only, and show that near the brane the contribution induced by the latter is larger than the one induced by the string; however, for points distant from the brane the situation is reversed. Moreover, some asymptotic expressions for the VEV of the field squared and EMT are provided for specific limiting cases of the physical parameters of the model. Also, an application of our results is given for a cosmic string in the $Z_2$-symmetric Randall-Sundrum braneworld model with a single brane.
\end{abstract}

\bigskip


\bigskip

\section{Introduction}
The anti-de Sitter (AdS) spacetime is one of the most interesting spacetimes allowed by the Theory of General Relativity.
Due to its maximal symmetry, many problems involving quantum fields propagating can be exactly solvable   
(see, for example \cite{Fronsdal}-\cite{Caldareli}). This allows to reveal information on the influence of gravitational field on quantum matter  in less symmetric geometries. In addition, the length scale related to the AdS negative constant curvature, can serve as a regularization parameter for infrared divergences in interacting quantum field theories without 
have to reduce the number of symmetries \cite{Callan1990}.  Besides, the importance of this theoretical background increased when it was discovered that AdS spacetime generically arises as a ground state in extended supergravity and in string theories. Additional interest in this subject was generated by the appearance of two models where AdS geometry plays a special role. The first model, the AdS/CFT correspondence (for a review see \cite{Ahar00}), represents a realization of the holographic principle and relates string theories or supergravity in the AdS bulk with a conformal field theory
living on its boundary. The second model is a realization of a braneworld scenario with large extra dimensions and
provides a solution to the hierarchy problem between the gravitational and 	electroweak mass scales (for reviews on braneworld gravity and cosmology see \cite{Brax03,Maar10}).

According to the Big Bang Theory, at the beginning the Universe was very hot and was in a complete symmetric stage. During its expansion, it cooled and underwent several phase changes, accompanied by spontaneous symmetry breaking resulting in the formation of toplological defects \cite{Kibble,V-S}. These include domain walls, cosmic strings and monopoles. Among them the cosmic strings are of special interest. 

Cosmic strings are linear topological defects. The gravitational field produced by an idealized cosmic string may be approximated by a planar angle deficit in the two-dimensional sub-space orthogonal to the string.  Although there is no Newtonian potential, the lack of global flatness is responsible for many interesting phenomena as shown many years ago by Linet \cite{Linet} and Smith \cite{Smith}. For instance, in Refs.\cite{Aliev1997,Aliev:1997fbq} the authors have studied the vacuum polarization effects induced by multiple parallel static straight-line cosmic strings and have shown that two parallel strings mutually attract each other with a Casimir-like force. Moreover, the presence of  the string allows effects such as  particle-antiparticle pair production by a single photon and bremsstrahlung radiation from charged particles which are not possible in empty Minkowski space due to the conservation of linear momentum \cite{Skarzhinsky}. The dimensionless parameter that characterizes the strength of gravitational interactions of strings with matter is its {\it tension}, that is given in natural units by $G\mu_0$, being $G$ the Newton's constant and $\mu_0$ its linear mass density, proportional to the square of the symmetry breaking scale energy.

In the eighties and early nineties of the last century, cosmic string was considered as possible seeds for large scale
structure formation in the Universe. Although recent observational data on the temperature anisotropies of the cosmic microwave background radiation (CMB) have excluded  the cosmic strings as the main origin of
structures, they are still sources for a number of interesting physical effects such as gamma ray bursts \cite{Berezinski}, gravitational waves \cite{Damour} and high energy cosmic rays \cite{Bhattacharjee}. Recently, cosmic strings have attracted renewed interest
partly because a variant of their formation mechanism is proposed in the framework of brane inflation \cite{Sarangi}-\cite{Dvali}.

The analysis of the VEV of the bosonic current density, $\langle j^\mu\rangle$, and the energy-momentum tensor, $\langle T^\mu_\nu\rangle$, induced by a  magnetic flux running along the core of an idealized cosmic string in a high-dimensional AdS spacetime, admitting that an extra dimension coordinate is compactified to a circle, were analyzed in \cite{Wagner_19} and \cite{Wagner_20}, respectively.  In both papers it was admitted the presence of an extra magnetic flux enclosed by the compactified dimension. Moreover, the analysis of VEV of fermionic current density and energy-momentum tensor in $(1+4)-$dimensional AdS spacetime in the presence of a cosmic string, considering the compactification of the extra dimension, have been developed in \cite{Wagner_20a} and \cite{Wagner_22}, respectively. Finally considering the presence of a brane parallel to the AdS boundary, the analysis of the effects of the brane on the vacuum fermionic current, $\langle j^\mu\rangle$, and the energy-momentum tensor, $\langle T^\mu_\nu\rangle$, were investigate in \cite{Wagner_21} and \cite{Wagner_22a}, respectively.

The vacuum polarization effects induced by a cosmic string in AdS spacetime were studied in Ref.\cite{deMello:2011ji}, where the authors have analyzed the VEVs of the field squared and the energy-momentum tensor. Here in this paper, we want to continue in this line of investigation and study the VEVs of the field squared and the energy-momentum tensor induced by an idealized cosmic string carrying a magnetic flux running along its core on $(1+D)$-dimensional AdS bulk considering the presence of a brane parallel to the AdS boundary. This analysis is developed for both part of the space. Moreover, we admit that the bosonic field obeys the Robin boundary condition (BC) on the brane in both parts of the space.

The paper is organized as follows. In the section \ref{sec2} we present the setup of the problem that we want investigate, and the complete set of normalized positive and negative energy solutions to the Klein-Gordon equation in the presence of a brane parallel to the AdS boundary. In the section \ref{sec3} we construct the Wightman function for both regions of the space. In the sections \ref{sec4} and \ref{sec5}, respectively, the VEVs of the field squared and the energy-momentum tensor in the region between the brane and the AdS horizon (R(right)-region) and the AdS boundary and the brane (L(left)-region) are investigated. Various asymptotic limits are considered and numerical results are presented. In the section \ref{sec6} we apply  our analysis to the Randall-Sundrum type model with a single brane and section \ref{sec:Conc} summarizes the most relevant results obtained. Throughout the paper, we use natural units $G=\hbar =c=1$.
\section{Model setup}
\label{sec2}
In this section we present the model setup, describing the background geometry and the matter field content. We begin by presenting the line element, in cylindrical coordinates, associated with the spacetime geometry we are going to consider, which is a $(D+1)$-dimensional anti-de Sitter (AdS) spacetime containing a cosmic string: 
\begin{equation}
	ds^{2}=g_{\mu\nu}dx^{\mu}dx^{\nu}=e^{-2y/a}\left[dt^{2}-dr^2-r^2d\phi^2- \sum_{i=4}^{D}(dx^{i})^2\right]-dy^2
	\ ,
	\label{ds1}
\end{equation}
where $r\geqslant 0$ and $\phi \in \lbrack 0,\ 2\pi /q]$ define the
coordinates on the conical geometry, $(t, z, y, x^i)\in (-\infty ,\ \infty )$ for $i=4,...,D$ and
the parameter $a$ determines the curvature scale of the background
spacetime. The latter is related to the cosmological constant, $\Lambda=-D(D-1)/(2a^{2})$, and the Ricci scalar, $R=-D(D+1)/a^{2}$. In the case of $D=3$, the cosmic string is assumed to be along the $y$-axis. Moreover, the presence of the cosmic string is codified through the parameter $q\geq 1$. Using the Poincarè coordinate defined by $w = ae^{y/a}$, the line element above can be conformally related to the line element associated with a cosmic string in Minkowski spacetime
\begin{equation}
	\label{HDCS}
	ds^2 = \left(\frac{a}{w}\right)^2\bigg[dt^2 - dr^2 - r^2d\phi^2 - dw^2 - \sum_{i=4}^{D}(dx^i)^2\bigg]   \   .
\end{equation}
For the new coordinate one has $w\in \lbrack 0,\ \infty )$. In particular, the values $w=0$ and $w=\infty $ correspond to the AdS boundary and horizon, respectively.

As to the matter field content, we will consider a charged massive bosonic field coupled to a gauge field, $A_{\mu }$. The corresponding field equation that governs the quantum dynamics is given by the Klein-Gordon (KG) equation,
\begin{equation}
	({\mathcal{D}}^2 + m^2 + \xi R)\varphi(x) = 0  \   , 
	\label{KGE}
\end{equation}
where the differential operator in the field equation reads
\begin{align}
	{\mathcal{D}}^2=\frac{1}{\sqrt{|g|}}{\mathcal{D}}_{\mu}\left(\sqrt{|g|}g^{\mu \nu} {\mathcal{D}}_{\nu
	}\right), \ {\mathcal{D}}_{\mu
	}=\partial _{\mu }+ieA_{\mu }\   \ {\rm with} \  \  g=\det(g_{\mu\nu})  \  .  \label{1}
\end{align}
In addition, we also consider the presence of a non-minimal coupling, $\xi$, between the field and the geometry represented by the Ricci scalar, $R$. Two specific values for the curvature coupling are of special interest: $\xi = 0$ and $\xi = \frac{D - 1}{4D}$, that correspond to minimal and conformal coupling, respectively. As to the vector potential, we consider the configuration $A_{\mu}=\delta_{\mu}^{\phi}A_{\phi}$ with $A_{\phi}$ constant, corresponding to a thin magnetic flux along the string's core.

We also consider a codimension one flat boundary, hereafter named brane, located at $w=w_0$ and parallel to the AdS boundary. On the brane we will impose that the field operator obeys the gauge invariant Robin boundary condition,
\begin{equation}
	(1+\beta n^{\mu}{\mathcal{D}}_{\mu
	})\varphi(x)=0, \ w=w_0 \ .
	\label{RBC}
\end{equation}
The inward pointing vector (with respect to the region under consideration), $n^{\mu}$, is normal to the brane at $w=w_0$. It is defined by $n^{\mu}=\delta_{(\rm{J})}\delta_{3}^{\mu}a/w$, where $\rm{J}=\rm{L}$, $\delta_{(\rm{L})}=-1$ in the region $0\le w\le w_0$, L-region,  and $\rm{J}=\rm{R}$, $\delta_{(\rm{R})}=1$ in the region $w_0\le w\le \infty$, R-region.
In addition, the parameter $\beta$ in \eqref{RBC} is a constant and it encodes the properties of the brane, which in the special cases $\beta=0$ and $\beta=\infty$ correspond to the Dirichlet and Neumann boundary conditions, respectively. Moreover, note that the value of this parameter for both the regions divided by the brane could be different in general.

In the geometry defined by \eqref{HDCS} and in the presence of the vector potential $A_{\mu}=\delta_{\mu}^{\phi}A_{\phi}$, the KG equation \eqref{KGE} becomes
\begin{eqnarray}
	\left[\frac{\partial^2}{\partial t^2} - \frac{\partial^2}{\partial r^2} - \frac{1}{r}\frac{\partial}{\partial r} - \frac{1}{r^2}\left(\frac{\partial}{\partial\phi} + ieA_{\phi}\right)^2
	- \frac{\partial^2}{\partial w^2}-\frac{(1-D)}{w}\frac{\partial}{\partial w}
	\right.\nonumber\\
	\left. + \frac{M(D,m,\xi)}{w^2} - \sum_{i=4}^{D}\frac{\partial^2}{\partial (x^i)^2} \right]\varphi(x) = 0  \  , 
	\label{KGE2}
\end{eqnarray}
where $M(D,m,\xi) = a^2m^2 - \xi D(D+1)$. 

According to the symmetry of the problem, the normalized positive energy wave function solutions of \eqref{KGE2} reads
\begin{equation}
	\varphi_{\sigma}(x) = C_{\sigma}w^{\frac{D}{2}}Z_{\nu}(pw)J_{q|n +\alpha|}(\lambda r)e^{-iE t + iqn\phi + i\vec{k}\cdot\vec{x}_{\parallel}} \ ,
	\label{Solu1}
\end{equation}
where we have defined the function
\begin{equation}
	Z_{\nu}(pw)=C_1J_{\nu}(pw)+C_2Y_{\nu}(pw)
	\label{W-function}
\end{equation}
is a linear combination of the Bessel and Neumann functions \cite{Abra}, with the order given by
\begin{equation}
	\nu = \sqrt{\frac{D^2}{4} + a^2m^2 - \xi D(D+1)} \ .
	\label{order}
\end{equation}
Moreover,
\begin{eqnarray}
	E &=& \sqrt{\lambda^2 + p^2 + \vec{k}^2},\nonumber\\
	\alpha &=& \frac{eA_{\phi}}{q} = -\frac{\Phi_{\phi}}{\Phi_0} \ , 
	\label{const}
\end{eqnarray}
being $\Phi_0=\frac{2\pi}{e}$, the quantum flux. In \eqref{Solu1} $\vec{x}_{\parallel}$ represents the coordinates defined in the $(D-4)$ extra dimensions, being $\vec{k}$ the corresponding momentum, and  $\sigma$ represents the set of quantum numbers $(n, \lambda, p, \vec{k})$, being $n=0,\pm1,\pm2,\ldots$, $\lambda \geq 0$, $-\infty<k^j<\infty$ for $j=4,...,D$. The quantum number $p$ is determined separately in each region divided by the brane.

The coefficient $C_{\sigma}$ in \eqref{Solu1} is determined from the normalization condition
\begin{eqnarray}
	\int d^Dx\sqrt{|g|}g^{00}\varphi_{\sigma'}^{*}(x)\varphi_{\sigma}(x)= \frac{1}{2E}\delta_{\sigma,\sigma'}  \   ,
	\label{NC}
\end{eqnarray}
where the delta symbol on the right-hand side is understood as Dirac delta 
function for the continuous quantum numbers, $\lambda$, $p$ and ${\vec{k}}$, and
Kronecker delta for the discrete one, $n$.

Let us first consider the R-region. By imposing the Robin boundary condition \eqref{RBC} on the flat boundary at $w=w_0$, we get the relation $C_2/C_1=-\bar{J}_{\nu}(pw_0)/\bar{Y}_{\nu}(pw_0)$ for the coefficients in \eqref{W-function}. Here and bellow we use the notation
\begin{equation}
	\bar{F}(x)=A_{0}F(x)+B_{0}xF^{\prime}(x) \ ,
\end{equation}
with the coefficients
\begin{equation}
	A_{0}=1+\delta_{(\rm{J})}\frac{D\beta}{2a}, \quad B_{0}=\delta_{(\rm{J})}\frac{\beta}{a} \ . 
\end{equation}
Thus, the mode functions in the R-region that obey the boundary condition \eqref{RBC} can be written presented as,
\begin{equation}
	\varphi_{(R)\sigma}(x) = C_{(R)\sigma}w^{\frac{D}{2}}g_{\nu}(pw_0,pw)J_{q|n +\alpha|}(\lambda r)e^{-iE t + iqn\phi + i\vec{k}\cdot\vec{x}_{\parallel}} \ ,
		\label{Solu-R}
\end{equation}
where we have introduced the function
\begin{equation}
	g_{\nu}(u,v)=J_{\nu}(v)\bar{Y}_{\nu}(u)-\bar{J}_{\nu}(u)Y_{\nu}(v) \ .
	\label{g-function}
\end{equation}
Taking into account the continuous spectrum of the quantum number $p$ and the normalization condition \eqref{NC}, we obtain
\begin{equation}
	|C_{(R)\sigma}|^2=\frac{(2\pi)^{2-D}qp\lambda}{2Ea^{D-1}[\bar{J}_{\nu}^2(pw_0)+\bar{Y}_{\nu}^2(pw_0)]} \ .
\end{equation}

In the L-region, the region of integration over $w$ in the normalization condition \eqref{NC} goes over the interval $0\le w\le w_0$. For the solutions with $C_2\neq0$ in \eqref{W-function}, the integral over $w$ diverges at the lower limit $w=0$ in the range of values $\nu\ge1$. Therefore, for this case we should take $C_2=0$ according to the normalization condition. On the other hand, in the region $0\le \nu <1$, the solution \eqref{W-function} with $C_2\neq0$ is normalizable and in
order to uniquely define the mode functions an additional boundary condition at the AdS boundary
is required \cite{Breitenlohner,Avis1978}. Here, we
will choose the Dirichlet condition which gives $C_2=0$. Thus, with this choice, the mode function in the L-region are given by
\begin{equation}
	\varphi_{(L)\sigma}(x) = C_{(L)\sigma}w^{\frac{D}{2}}J_{\nu}(pw)J_{q|n +\alpha|}(\lambda r)e^{-iE t + iqn\phi + i\vec{k}\cdot\vec{x}_{\parallel}} \ ,
	\label{Solu-L}
\end{equation}
where, according to the Robin boundary condition \eqref{RBC} on the brane, the eigenvalues of the quantum number $p$ obey the relation:
\begin{equation}
	\bar{J}_{\nu}(pw_0)=0 \ ,
\end{equation}
where the eigenvalues are given by $p=p_{\nu,i}/w_0$, with $p_{\nu,i}$ being the positive zeros of the function $\bar{J}_{\nu}(x)$, enumerated by $i=1, 2,...$. Note that the roots $p_{\nu,i}$ do not depend on the location of the brane. From the normalization condition \eqref{NC}, with $\delta_{p,p^\prime}=\delta_{i,i^\prime}$, and integrating over $w$ in the interval $[0,w_0]$, we get
\begin{equation}
	|C_{(L)\sigma}|^2=\frac{(2\pi)^{2-D}qp_{\nu,i}\lambda T_{\nu}(p_{\nu,i})}{w_0a^{D-1}\sqrt{p_{\nu,i}^2+w_{0}^2(\lambda^2+\vec{k}^2})}\ ,
\end{equation}
with the function $T_{\nu}(z)=z[(z^2-\nu^2)J_{\nu}^2(z)+z^2(J_{\nu}^{\prime}(z))^2]^{-1}$.
\section{Wightman Function}\label{sec3}
In this section we present the positive frequency Wightman function, $W(x,x^{\prime})=\langle 0| \hat{\varphi}(x)\hat{\varphi}^{\dagger}(x^{\prime})|0\rangle$, where $|0\rangle$ stands for the vacuum state, for both L-region and R-region in a closed form. Here we will assume that the field is prepared in the Poincaré vacuum state. To evaluate this function, we use the mode sum formula:
\begin{equation}
	W(x,x^{\prime})=\sum_{\sigma}\varphi_{\sigma}(x)\varphi_{\sigma}^{\ast}(x^{\prime}) \ .
	\label{Wightman-function}
\end{equation}
\subsection{R-region}
Let us start with the R-region by taking the respective wave function solutions \eqref{Solu-R} into the above expression. Thus, we have
\begin{eqnarray}
	W(x,x^\prime)&=&\frac{q(ww^\prime)^{D/2}}{2(2\pi)^{D-2}a^{D-1}}\sum_{\sigma}\frac{p\lambda}{E}\frac{g_{\nu}(pw_0,pw)g_{\nu}(pw_0,pw^{\prime})}{\bar{J}_{\nu}^2(pw_0)+\bar{Y}_{\nu}^2(pw_0)}J_{q|n+\alpha|}(\lambda r)J_{q|n+\alpha|}(\lambda r^\prime)\nonumber \\
	&\times&e^{iqn\Delta\phi+i\vec{k}\cdot \Delta\vec{x}_{\parallel}-iE\Delta t} \ ,
\end{eqnarray}
	where $g_{\nu}(u,v)$ is defined in \eqref{g-function}, $\Delta t=t-t^\prime$, $\Delta\phi=\phi-\phi^\prime$, $\Delta \vec{x}_{\parallel}=\vec{x}_{\parallel}-\vec{x}_{\parallel}^\prime$ and with the notation
\begin{equation}
	\sum_{\sigma}=\int_{0}^{\infty}d\lambda\int_{0}^{\infty}dp\sum_{n}\int d\vec{k} \ .
\end{equation}
Now performing a Wick rotation on the time coordinate and using the identity
\begin{equation}
	\frac{e^{-E\Delta \tau}}{E}=\frac2{\sqrt{\pi}}\int_0^\infty ds e^{-s^2E^2-\Delta \tau^2/(4s^2)}  \  ,
	\label{identity}
\end{equation}
with the energy given by $E=\sqrt{\lambda^2+p^2+\vec{k}^2}$ in the R-region, we can integrate over $\lambda$ and $\vec{k}$ with the help of \cite{Grad}. The result is,
\begin{eqnarray}
	W(x,x^\prime)&=&\frac{qrr^\prime}{2(2\pi)^{D/2}a^{D-1}}\left(\frac{ww^\prime}{rr^\prime}\right)^{D/2}\int_{0}^{\infty}d\chi\chi^{\frac{D}{2}-2}e^{-\frac{r^2+r^{\prime2}+\Delta\vec{x}_{\parallel}^2-\Delta t^2}{2rr^\prime}\chi}\sum_{n}e^{inq\Delta\phi}I_{q|n+\alpha|}(\chi)\nonumber\\
	&\times&\int_{0}^{\infty}dppe^{-\frac{rr^\prime}{2\chi}p^2}\frac{g_{\nu}(pw_0,pw)g_{\nu}(pw_0,pw^{\prime})}{\bar{J}_{\nu}^2(pw_0)+\bar{Y}_{\nu}^2(pw_0)} \ ,
	\label{W-function_full}
\end{eqnarray}
where we have introduced a new variable, $\chi=rr^\prime/(2s^2)$.

Note that the Wightman function above presents the contributions coming from the cosmic string and the boundary. However, in this paper we are mainly interested to investigate the vacuum polarization effects associated to the boundary. In this sense, we will split those contributions and focus on the boundary induced one. For this end, we proceed in the following way:
\begin{equation}
	W_{b}(x,x^\prime)=W(x,x^\prime)-W_{cs}(x,x^\prime) \ ,
	\label{Proc}
\end{equation}
where the term induced by the cosmic string was calculated in \cite{Wagner_19} and is given by
\begin{eqnarray}
	W_{cs}(x,x^\prime)&=&\frac{qrr^\prime}{2(2\pi)^{D/2}a^{D-1}}\left(\frac{ww^\prime}{rr^\prime}\right)^{D/2}\int_{0}^{\infty}d\chi\chi^{\frac{D}{2}-2}e^{-\frac{r^2+r^{\prime2}+\Delta\vec{x}_{\parallel}^2-\Delta t^2}{2rr^\prime}\chi}\sum_{n}e^{inq\Delta\phi}I_{q|n+\alpha|}(\chi)\nonumber\\
	&\times&\int_{0}^{\infty}dppe^{-\frac{rr^\prime}{2\chi}p^2}J_{\nu}(pw)J_{\nu}(pw^\prime) \ .
	\label{W-function_cs}
\end{eqnarray}
Thus, replacing \eqref{W-function_full} and \eqref{W-function_cs} into \eqref{Proc} and using the identity \cite{Mello2006}
\begin{eqnarray}
	&&\frac{g_{\nu}(pw_0,pw)g_{\nu}(pw_0,pw^{\prime})}{\bar{J}_{\nu}^2(pw_0)+\bar{Y}_{\nu}^2(pw_0)}-J_{\nu}(pw)J_{\nu}(pw^\prime)\nonumber\\
	&=&-\frac{1}{2}\sum_{l=1}^{2}\frac{\bar{J}_{\nu}(pw_0)}{\bar{H}_{\nu}^{(l)}(pw_0)}H_{\nu}^{(l)}(pw)H_{\nu}^{(l)}(pw^\prime) \ ,
\end{eqnarray}
we get
\begin{eqnarray}
	W_{b}(x,x^\prime)&=&-\frac{qrr^\prime}{4(2\pi)^{D/2}a^{D-1}}\left(\frac{ww^\prime}{rr^\prime}\right)^{D/2}\int_{0}^{\infty}d\chi\chi^{\frac{D}{2}-2}e^{-\frac{r^2+r^{\prime2}+\Delta\vec{x}_{\parallel}^2-\Delta t^2}{2rr^\prime}\chi}\sum_{n}e^{inq\Delta\phi}I_{q|n+\alpha|}(\chi)\nonumber\\
	&\times&\int_{0}^{\infty}dppe^{-\frac{rr^\prime}{2\chi}p^2}\sum_{l=1}^{2}\frac{\bar{J}_{\nu}(pw_0)}{\bar{H}_{\nu}^{(l)}(pw_0)}H_{\nu}^{(l)}(pw)H_{\nu}^{(l)}(pw^\prime) \ ,
	\label{W-function_b}
\end{eqnarray}
where $H_{\nu}^{(l)}(x)$, $l=1, 2$, are the Hankel functions \cite{Abra}.

The parameter $\alpha$ in Eq.\eqref{const} can be written in the form
\begin{equation}
	\alpha=n_{0}+\alpha_0, \ \textrm{with}\ |\alpha_0|<\frac{1}{2},
	\label{const-2}
\end{equation}
being $n_{0}$ an integer number. This allow us to sum over the quantum number $n$ in \eqref{W-function_b}, using the result obtained in \cite{deMello:2014ksa}, given below,
\begin{eqnarray}
	&&\sum_{n=-\infty}^{\infty}e^{iqn\Delta\phi}I_{q|n+\alpha|}(\chi)=\frac{1}{q}\sum_{k}e^{\chi\cos(2\pi k/q-\Delta\phi)}e^{i\alpha(2\pi k -q\Delta\phi)}\nonumber\\
	&-&\frac{e^{-iqn_{0}\Delta\phi}}{2\pi i}\sum_{j=\pm1}je^{ji\pi q|\alpha_0|}
	\int_{0}^{\infty}dy\frac{\cosh{[qy(1-|\alpha_0|)]}-\cosh{(|\alpha_0| qy)e^{-iq(\Delta\phi+j\pi)}}}{e^{\chi\cosh{(y)}}\big[\cosh{(qy)}-\cos{(q(\Delta\phi+j\pi))}\big]} \ ,
	\label{summation-formula}
\end{eqnarray}
where
\begin{equation}
	-\frac{q}{2}+\frac{\Delta\phi}{\Phi_{0}}\le k\le \frac{q}{2}+\frac{\Delta\phi}{\Phi_{0}}  \   .
\end{equation}
Substituting the formula above into \eqref{W-function_b}, we can perform the integration over $\chi$ using the integral formula \cite{Grad}
\begin{equation}
	\int_{0}^{\infty} x^{\nu-1}e^{-\frac{\beta}{x}-\gamma x}dx=2\left(\frac{\beta}{\gamma}\right)^{\nu/2}K_{\nu}(2\sqrt{\beta\gamma}) \ .
	\label{integral-formula}
\end{equation}
The result of these operations is the following:
\begin{eqnarray}
	W_{b}(x,x^\prime)&=&-\frac{(ww^\prime)^{D/2}}{2(2\pi)^{D/2}a^{D-1}}\Biggl\{\sum_{k}\frac{e^{i\alpha(2\pi k -q\Delta\phi)}}{u_k^{\frac{D}{2}-1}}\int_{0}^{\infty}dpp^{D/2}\sum_{l=1}^{2}\frac{\bar{J}_{\nu}(pw_0)}{\bar{H}_{\nu}^{(l)}(pw_0)}\nonumber\\
	&\times&H_{\nu}^{(l)}(pw)H_{\nu}^{(l)}(pw^\prime)K_{\frac{D}{2}-1}(pu_k)-\frac{qe^{-iqn_{0}\Delta\phi}}{2\pi i}\sum_{j=\pm1}je^{ji\pi q|\alpha_0|}
	\nonumber\\
	&\times&\int_{0}^{\infty}dy\frac{\cosh{[qy(1-|\alpha_0|)]}-\cosh{(|\alpha_0| qy)e^{-iq(\Delta\phi+j\pi)}}}{u_y^{\frac{D}{2}-1}\big[\cosh{(qy)}-\cos{(q(\Delta\phi+j\pi))}\big]}\int_{0}^{\infty}dpp^{D/2}\sum_{l=1}^{2}\frac{\bar{J}_{\nu}(pw_0)}{\bar{H}_{\nu}^{(l)}(pw_0)}\nonumber\\
	&\times&H_{\nu}^{(l)}(pw)H_{\nu}^{(l)}(pw^\prime)K_{\frac{D}{2}-1}(pu_y)\Biggr\} \ ,
	\label{W-function_b-2}
\end{eqnarray}
where we have introduced the notation
\begin{eqnarray}
	u_{k}^2&=&r^2+r'^2-2rr'\cos{(2\pi k/q-\Delta\phi)}
	+\Delta\vec{x}_{\parallel}^{2}-\Delta t^2\nonumber\\
	u_{y}^2&=&r^2+r'^2+2rr'\cosh{(y)}+\Delta\vec{x}_{\parallel}^{2}-\Delta t^2.
	\label{arg}
\end{eqnarray}
As the last step, we rotate the contour integration over $p$ by the angle $\pi/2$ $(-\pi/2)$ for the term $l=1$ $(l=2)$. The result is
\begin{eqnarray}
	W_{b}(x,x^\prime)&=&-\frac{(ww^\prime)^{D/2}}{(2\pi)^{D/2}a^{D-1}}\int_{0}^{\infty}dpp^{D-1}\frac{\bar{I}_{\nu}(pw_0)}{\bar{K}_{\nu}(pw_0)}K_{\nu}(pw)K_{\nu}(pw^\prime)\nonumber\\
	&\times&\Biggl\{\sum_{k}e^{i\alpha(2\pi k -q\Delta\phi)}f_{\frac{D}{2}-1}(pu_k)-\frac{qe^{-iqn_{0}\Delta\phi}}{2\pi i}\sum_{j=\pm1}je^{ji\pi q|\alpha_0|}
	\nonumber\\
	&\times&\int_{0}^{\infty}dy\frac{\cosh{[qy(1-|\alpha_0|)]}-\cosh{(|\alpha_0| qy)e^{-iq(\Delta\phi+j\pi)}}}{\cosh{(qy)}-\cos{(q(\Delta\phi+j\pi))}}f_{\frac{D}{2}-1}(pu_y)\Biggl\} \ ,
	\label{W-function_b-3}
\end{eqnarray}
where we have introduced the notation
\begin{equation}
	f_{\mu}(x)=\frac{J_{\mu}(x)}{x^{\mu}} \ .
	\label{func-f}
\end{equation}
\subsection{L-region}
Now we want to calculate the Wightman function in the L-region. Taking the respective wave function solutions \eqref{Solu-L} into \eqref{Wightman-function}, we get
\begin{eqnarray}
	W(x,x^\prime)&=&\frac{q(ww^\prime)^{D/2}}{(2\pi)^{D-2}a^{D-1}w_0^2}\sum_{\sigma}\frac{\lambda p_{\nu,i}}{\sqrt{(p_{\nu,i}/w_0)^2+\lambda^2+\vec{k}^2}}T_{\nu}(p_{\nu,i})J_{\nu}(p_{\nu,i}w/w_0)J_{\nu}(p_{\nu,i}w^\prime/w_0)
	\nonumber\\
	&\times&J_{q|n+\alpha|}(\lambda r)J_{q|n+\alpha|}(\lambda r^\prime)e^{inq\Delta\phi+i\vec{k}\cdot\Delta \vec{x}_{\parallel}-iE\Delta t} \ ,
\end{eqnarray}
with the notation
\begin{equation}
	\sum_{\sigma}=\int_{0}^{\infty}d\lambda\sum_{i=1}^{\infty}\sum_{n}\int d\vec{k} \ .
\end{equation}
Once again making a Wick rotation on the time coordinate and using the identity \eqref{identity} with the energy in the L-region given by $E=\sqrt{\lambda^2+\vec{k}^2+(p_{\nu,i}/w_0)^2}$, we can integrate over $\lambda$ and $\vec{k}$, obtaining the following result:
\begin{eqnarray}
	W(x,x^\prime)&=&\frac{qrr^\prime}{(2\pi)^{D/2}a^{D-1}w_0^2}\left(\frac{ww^\prime}{rr^\prime}\right)^{D/2}\int_{0}^{\infty}d\chi\chi^{\frac{D}{2}-2}e^{-\frac{r^2+r^{\prime2}+\Delta\vec{x}_{\parallel}^2-\Delta t^2}{2rr^\prime}\chi}\sum_{n}e^{inq\Delta\phi}I_{q|n+\alpha|}(\chi)\nonumber\\
	&\times&\sum_{i=1}^{\infty}p_{\nu,i}e^{-\frac{rr^\prime}{2\chi w_0^2}p_{\nu,i}^2}T_{\nu}(p_{\nu,i})J_{\nu}(p_{\nu,i}w/w_0)J_{\nu}(p_{\nu,i}w^\prime/w_0) \ ,
	\label{W-function_full-L}
\end{eqnarray}
where we have introduced the variable $\chi=rr^\prime/(2s^2)$. Now, writing the parameter $\alpha$ as in \eqref{const-2} once again and using the formula \eqref{summation-formula} for the summation over $n$, we can perform the integration over $\chi$ using the formula \eqref{integral-formula}. Following these steps, we get the expression
\begin{eqnarray}
	W(x,x^\prime)&=&\frac{2(ww^\prime)^{D/2}}{(2\pi)^{D/2}a^{D-1}w_0^{\frac{D}{2}+1}}\Biggl\{\sum_{k}\frac{e^{i\alpha(2\pi k -q\Delta\phi)}}{u_k^{\frac{D}{2}-1}}\sum_{i=1}^{\infty}p_{\nu,i}^{D/2}T_{\nu}(p_{\nu,i})J_{\nu}(p_{\nu,i}w/w_0)J_{\nu}(p_{\nu,i}w^\prime/w_0)\nonumber\\
	&\times&K_{\frac{D}{2}-1}(u_kp_{\nu,i}/w_0)-\frac{qe^{-iqn_{0}\Delta\phi}}{2\pi i}\sum_{j=\pm1}je^{ji\pi q|\alpha_0|}
	\nonumber\\
	&\times&\int_{0}^{\infty}dy\frac{\cosh{[qy(1-|\alpha_0|)]}-\cosh{(|\alpha_0| qy)e^{-iq(\Delta\phi+j\pi)}}}{u_y^{\frac{D}{2}-1}\big[\cosh{(qy)}-\cos{(q(\Delta\phi+j\pi))}\big]}\nonumber\\
	&\times&\sum_{i=1}^{\infty}p_{\nu,i}^{D/2}T_{\nu}(p_{\nu,i})J_{\nu}(p_{\nu,i}w/w_0)J_{\nu}(p_{\nu,i}w^\prime/w_0)K_{\frac{D}{2}-1}(u_yp_{\nu,i}/w_0)\Biggl\} \ ,
	\label{W-function_b-2-L}
\end{eqnarray}
where the variable $u_k$ and $u_y$ are defined in \eqref{arg}.

In order to develop the summation over $i$, we will use the generalized Abel-Plana summation formula \cite{SahaRev}
\begin{equation}
	\sum_{i=1}^{\infty}T_{\nu}(p_{\nu,i})f(p_{\nu,i})=\frac{1}{2}\int_{0}^{\infty}dzf(z)-\frac{1}{2\pi}\int_{0}^{\infty}dz\frac{\bar{K}_{\nu}(z)}{\bar{I}_{\nu}(z)}\left[e^{-i\nu z}f(iz)+e^{i\nu z}f(-iz)\right] \ .
	\label{Abel-Plana}
\end{equation}
For the problem under consideration, the function $f(z)$ is
\begin{equation}
	f(z)=z^{D/2}J_{\nu}(zw/w_0)J_{\nu}(zw^\prime/w_0)K_{\frac{D}{2}-1}(2uz/w_0) \ .
	\label{func}
\end{equation}
The first term in \eqref{Abel-Plana} will provide the Wightman function induced by the cosmic string in the absence of brane, while the second one is induced by the boundary. For the same argument given in the previous subsection, we concentrate on the brane-induced term. Therefore, after a few intermediate steps, we find
\begin{eqnarray}
	W_{b}(x,x^\prime)&=&-\frac{(ww^\prime)^{D/2}}{(2\pi)^{D/2}a^{D-1}}\int_{0}^{\infty}dvv^{D/2}\frac{\bar{K}_{\nu}(vw_0)}{\bar{I}_{\nu}(vw_0)}I_{\nu}(vw)I_{\nu}(vw^\prime)\nonumber\\
	&\times&\Biggl\{\sum_{k}e^{i\alpha(2\pi k -q\Delta\phi)}f_{\frac{D}{2}-1}(u_kv)-\frac{qe^{-iqn_{0}\Delta\phi}}{2\pi i}\sum_{j=\pm1}je^{ji\pi q|\alpha_0|}
	\nonumber\\
	&\times&\int_{0}^{\infty}dy\frac{\cosh{[qy(1-|\alpha_0|)]}-\cosh{(|\alpha_0| qy)e^{-iq(\Delta\phi+j\pi)}}}{\cosh{(qy)}-\cos{(q(\Delta\phi+j\pi))}}f_{\frac{D}{2}-1}(u_yv)\Biggl\} \ ,
	\label{W-function_b-3-L}
\end{eqnarray}
where we have made change of variable $z=vw_0$.
\section{Field Squared}\label{sec4}
The VEV of the field squared is formally obtained from the Wightman function by taking the coincidence limit, as shown below:
\begin{equation}
	\langle|\varphi(x)|^2\rangle=\lim_{x^\prime\rightarrow x}W(x,x^\prime) \ ,
	\label{FS-Formal}
\end{equation}
where the notation $|\varphi(x)|^2$ is understood here as $\varphi(x)\varphi^{\dagger}(x)$.
As we will see in what follows, in both regions, R and L, the VEV of the field squared can be decomposed in the form:
\begin{equation}
	\langle|\varphi(x)|^2\rangle_{b} = \langle|\varphi(x)|^2\rangle_{b}^{(0)} + \langle|\varphi(x)|^2\rangle_{b}^{(q,\alpha_0)} \ ,
	\label{decomp_phi2}
\end{equation}
where the first term in the right-hand side comes from the term $k=0$ in the sum over $k$ in the Wightman functions for both regions, \eqref{W-function_b-3} and \eqref{W-function_b-3-L}, and it is purely induced by the brane and was analyzed in \cite{Saharian:2003qs}, while the second term is a new contribution induced by the cosmic string and the magnetic flux along its core in the AdS geometry with a flat boundary. In this paper we are mainly interested in the investigation of the latter contribution and henceforth our study is restricted to it.
\subsection{R-region}
Taking the positive frequency Wightman function in the R-region, Eq. \eqref{W-function_b-3}, into the equation above and subsequently taking the coincidence limit, the second term in \eqref{decomp_phi2} reads
\begin{eqnarray}
	\langle|\varphi(x)|^2\rangle_{b}^{(q,\alpha_0)}&=&-\frac{2w^{D}}{(2\pi)^{D/2}a^{D-1}}\int_{0}^{\infty}dpp^{D-1}\frac{\bar{I}_{\nu}(pw_0)}{\bar{K}_{\nu}(pw_0)}K_{\nu}^2(pw)\Biggl[\sideset{}{'}\sum_{k=1}^{[q/2]}\cos(2\pi k\alpha_0)f_{\frac{D}{2}-1}(2rps_k)
	\nonumber\\
	&-&\frac{q}{2\pi }\int_{0}^{\infty}dy\frac{h(q,	\alpha_0,y)}{\cosh{(qy)}-\cos{(q\pi)}}f_{\frac{D}{2}-1}(2rp\cosh(y/2))\Biggl] \ ,
	\label{FS-R}
\end{eqnarray}
where $[q/2]$ represents the integer part of $q/2$, and the prime on the sign of the summation over $k$ means that for even values of $q$, the term $k=q/2$ should be
taken with the coefficient $1/2$. Moreover, henceforth we adopt the notation
\begin{equation}
	s_{k}=\sin(\pi k/q) \  .
\end{equation}
As to the function $h(q,\alpha_0,y)$, it reads
\begin{equation}
	h(q,\alpha_0,y)=\cosh(qy(1-|\alpha_0|))\sin(q\pi|\alpha_0|)+\cosh(qy|\alpha_0|)\sin(q\pi(1-|\alpha_0|)) \ .
\end{equation}
Moreover, note that by making a simple change of variable, $z=pw_0$, this VEV depends on the ratio $r/w$, which is related to the proper distance from the string, and the ratio $w/w_0$, which is related to the proper distance from the brane
\begin{equation}
	w/w_0=e^{(y-y_0)/a} \ .
\end{equation}
This feature is also present in the VEV of the squared field in the L-region and the components of the energy-momentum tensor as we will see below.

Let us investigate some special and asymptotic cases for the formula above. For a massless conformal scalar quantum field, we have $\nu=1/2$, according to \eqref{order}. Thus, using the corresponding modified Bessel functions for this order, we get
\begin{eqnarray}
	\langle|\varphi(x)|^2\rangle_{b}^{(q,\alpha_0)}&=&-\frac{2w^{D-1}}{(2\pi)^{D/2}a^{D-1}}\int_{0}^{\infty}dpp^{D-2}e^{-(2w-w_0)p}\frac{(2A_0-B_0) \sinh (pw_0)+2 B_0 pw_0 \cosh (pw_0)}{2A_0-B_0 (1+2pw_0)}\nonumber\\
	&\times&\Biggl[\sideset{}{'}\sum_{k=1}^{[q/2]}\cos(2\pi k\alpha_0)f_{\frac{D}{2}-1}(2rps_k)
	-\frac{q}{2\pi }\int_{0}^{\infty}dy\frac{h(q,	\alpha_0,y)}{\cosh{(qy)}-\cos{(q\pi)}}\nonumber\\
	&\times&f_{\frac{D}{2}-1}(2rp\cosh(y/2))\Biggl] \ .
	\label{FS_MSF}
\end{eqnarray}
Note that this result contrasts with the one for the pure brane-induced term, $\langle|\varphi|^2\rangle_{b}^{(0)}$ which is zero for a conformal massless quantum scalar field \cite{Saharian:2003qs}. 

For points with the proper distance from the plate much larger compared with the AdS radius, one has $w/w_0\gg1$. Introducing a new variable $u=pw$ and by making use of the formulae for the modified Bessel functions for small values of the argument \cite{Abra}, with the assumption that $A_0-\nu B_0\neq0$, to the leading order, we get
\begin{eqnarray}
	\langle|\varphi(x)|^2\rangle_{b}^{(q,\alpha_0)}&\approx&-\frac{2^{2-2\nu-D/2}}{\pi^{D/2}\Gamma(\nu)\Gamma(\nu+1) a^{D-1}}\left(\frac{A_0+\nu B_0}{A_0-\nu B_0}\right)\left(\frac{w_0}{w}\right)^{2\nu}\int_{0}^{\infty}duu^{D+2\nu-1}K_{\nu}^2(u)\nonumber\\
	&\times&\Biggl[\sideset{}{'}\sum_{k=1}^{[q/2]}\cos(2\pi k\alpha_0)f_{\frac{D}{2}-1}(2rus_k/w)
	-\frac{q}{2\pi }\int_{0}^{\infty}dy\frac{h(q,	\alpha_0,y)}{\cosh{(qy)}-\cos{(q\pi)}}\nonumber\\
	&\times&f_{\frac{D}{2}-1}(2ru\cosh(y/2)/w)\Biggl] \ .
	\label{FS-asymp-R}
\end{eqnarray}

Finally, we analyze the Minkowskian limit, $a\rightarrow\infty$, with fixed coordinate $y$. In this limit, the geometry under consideration is reduced to the geometry of a cosmic string in the background of a ($D+1$)-dimensional Minkowski spacetime. It can be observed that the coordinate $w$ in the arguments of the modified Bessel functions in this limit can be expressed as $w \approx a + y$. Considering that $\nu\gg1$, it can be seen that as we approach the Minkowskian limit, the order and the argument of the modified Bessel functions in equation \eqref{FS-R} tend to become large. Hence, we can make use of the corresponding uniform asymptotic expansions. Thus, for the leading order, one gets
\begin{eqnarray}
	\langle|\varphi(x)|^2\rangle_{b}^{(q,\alpha_0),(\rm{M})}&=&-\frac{1}{(2\pi)^{D/2}}\int_{m}^{\infty}du(u^2-m^2)^{\frac{D}{2}-1}\Biggl[\sideset{}{'}\sum_{k=1}^{[q/2]}\cos(2\pi k\alpha_0)\nonumber\\
	&\times&f_{\frac{D}{2}-1}(2rs_k\sqrt{u^2-m^2})
	-\frac{q}{2\pi }\int_{0}^{\infty}dy\frac{h(q,	\alpha_0,y)}{\cosh{(qy)}-\cos{(q\pi)}}\nonumber\\
	&\times&f_{\frac{D}{2}-1}(2r\cosh(y/2)\sqrt{u^2-m^2})\Biggl]\frac{1+\beta u}{1-\beta u}e^{-2u(y-y_0)} \ .
	\label{FS-R-M_case}
\end{eqnarray}

\begin{figure}[!htb]
	\begin{center}
		\centering
		\includegraphics[scale=0.3]{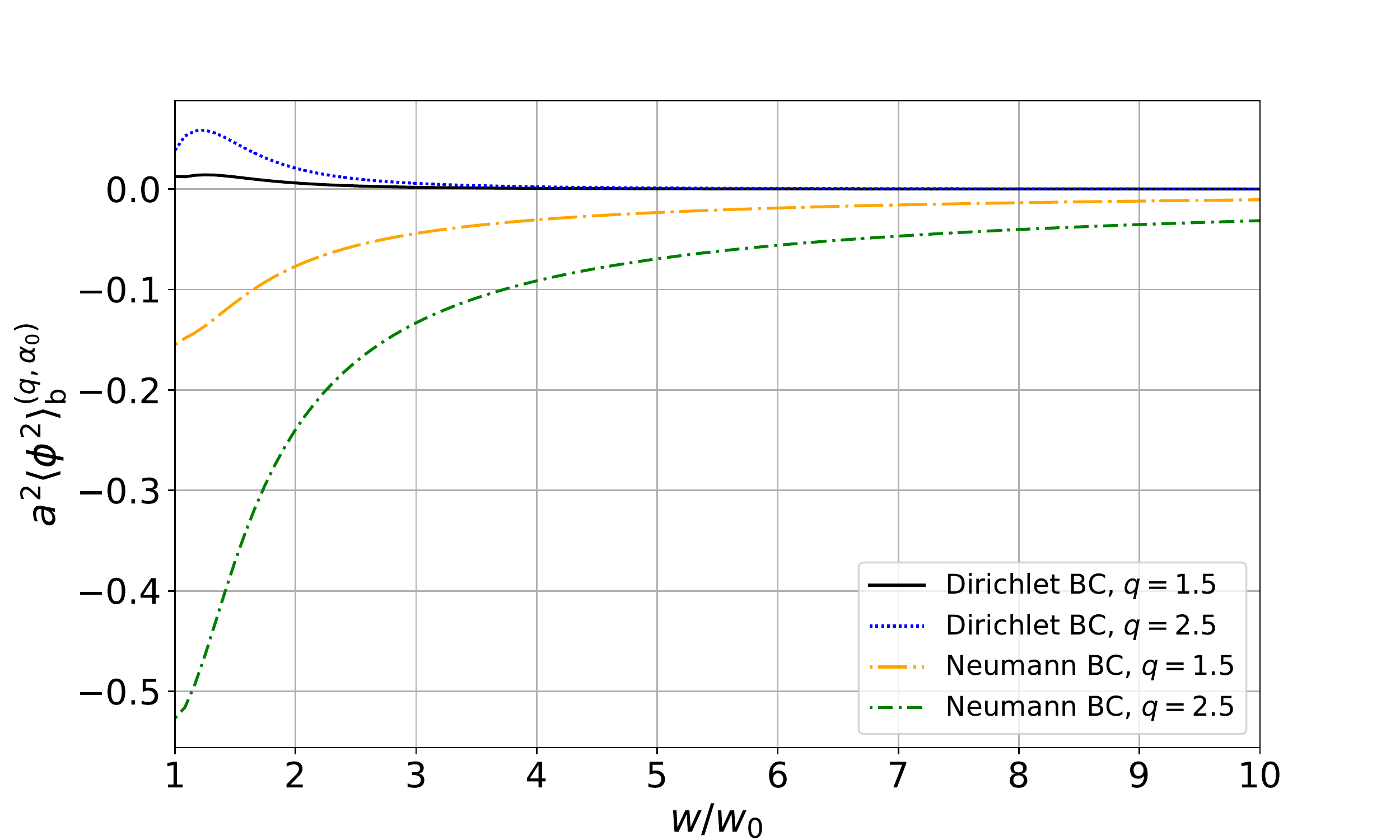}
		\quad
		\includegraphics[scale=0.3]{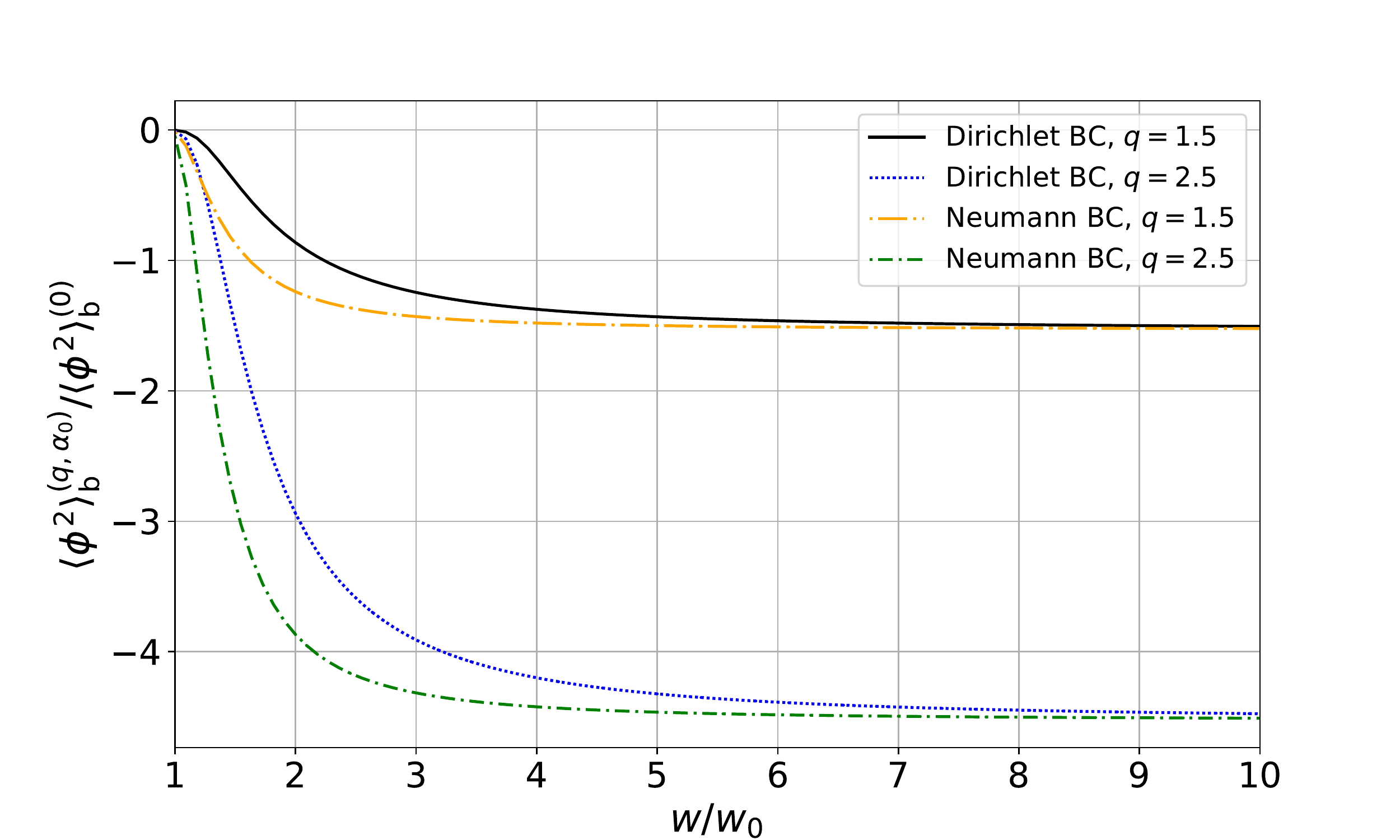}
		\caption{The VEV of the field squared $\langle|\varphi|^2\rangle_{b}^{(q,\alpha_0)}$ (left panel) and the ratio $\langle|\varphi|^2\rangle_{b}^{(q,\alpha_0)}/\langle|\varphi|^2\rangle_{b}^{(0)}$ (right panel) are exhibited as functions of $w/w_0$ for different values of $q$, considering Dirichlet and Neumann boundary conditions. The graphs are plotted for a minimally coupled massless scalar field in $D=3$ with fixed parameters $r/w_0=0.5$ and $\alpha_0=0.4$.}
		\label{fig1}
	\end{center}
\end{figure}

In Fig. \ref{fig1} we exhibit the dependence of the field squared, $\langle|\varphi|^2\rangle_{b}^{(q,\alpha_0)}$, (left panel) and the ratio  $\langle|\varphi|^2\rangle_{b}^{(q,\alpha_0)}/\langle|\varphi|^2\rangle_{b}^{(0)}$ (right panel) as functions of $w/w_0$ for Dirichlet and Neumann boundary conditions with different values of the deficit angle parameter, $q$. As we can see from the left panel the VEV of the field squared, induced by the cosmic string only, is finite on the brane and goes to zero for large distances from the brane with $(w_0/w)^{2\nu}$ according to the corresponding asymptotic formula \eqref{FS-asymp-R}. The right panel shows that the pure brane-induced contribution is dominant near the brane, while the string-induced term is more relevant for distant points from the brane. Moreover, note that the intensities of this VEV increase with the string parameter, $q$, and are higher for the Neumann BC.
\\
\subsection{L-region}
Now substituting the Wightman function in the L-region \eqref{W-function_b-3-L} into the formal expression for the field squared \eqref{FS-Formal} and taking the coincidence limit, the second term in the right hand side of \eqref{decomp_phi2} reads:
\begin{eqnarray}
	\langle|\varphi(x)|^2\rangle_{b}^{(q,\alpha_0)}&=&-\frac{2w^{D}}{(2\pi)^{D/2}a^{D-1}}\int_{0}^{\infty}dvv^{D-1}\frac{\bar{K}_{\nu}(vw_0)}{\bar{I}_{\nu}(vw_0)}I_{\nu}^2(vw)\Biggl[\sideset{}{'}\sum_{k=0}^{[q/2]}\cos(2\pi k\alpha_0)f_{\frac{D}{2}-1}(2rvs_k)
	\nonumber\\
	&-&\frac{q}{2\pi }\int_{0}^{\infty}dy\frac{h(q,	\alpha_0,y)}{\cosh{(qy)}-\cos{(q\pi)}}f_{\frac{D}{2}-1}(2rv\cosh(y/2))\Biggl] \ .
	\label{FS-L}
\end{eqnarray}
Comparing with \eqref{FS-R}, we see that the brane-induced contribution in the L-region is obtained from the corresponding quantity for the R-region by the replacements $I\rightarrow K$,
$K\rightarrow I$ of the modified Bessel functions.

Note that an important result which can be observed from the expression above and Eq. \eqref{FS-R} is that the VEVs of the field squared in both regions are finite on the brane. The corresponding values can be obtained directly from \eqref{FS-R} and \eqref{FS-L} by putting $w=w_0$. This finiteness of the field squared is in clear contrast to the behaviour of the pure brane-induced contribution, $\langle|\varphi(x)|^2\rangle_{b}^{(0)}$, which diverges on the brane as $1/(w-w_0)^{D-1} $\cite{Saharian:2003qs}.

As we have proceeded in R-region, let us now analyze some special and limiting cases of the squared field given above. For a massless conformal scalar quantum field in this region, we have
\begin{eqnarray}
	\langle|\varphi(x)|^2\rangle_{b}^{(q,\alpha_0)}&=&-\frac{2w^{D-1}}{(2\pi)^{D/2}a^{D-1}}\int_{0}^{\infty}dpp^{D-2}e^{-pw_0}\frac{ [2A_0-B_0(1+2pw_0)]\sinh^2(pw)}{(2 A_0-B_0) \sinh (pw_0)+2 B_0 pw_0 \cosh (pw_0)}\nonumber\\
	&\times&\Biggl[\sideset{}{'}\sum_{k=0}^{[q/2]}\cos(2\pi k\alpha_0)f_{\frac{D}{2}-1}(2rps_k)
	-\frac{q}{2\pi }\int_{0}^{\infty}dy\frac{h(q,	\alpha_0,y)}{\cosh{(qy)}-\cos{(q\pi)}}\nonumber\\
	&\times&f_{\frac{D}{2}-1}(2rp\cosh(y/2))\Biggl] \ .
	\label{FS-MSF-L}
\end{eqnarray}

We now analyse the asymptotic behaviour for points near the AdS boundary (hyperplane at $w=0$), $w\ll w_0$, with the proper distances from the plane much larger compared with the AdS curvature radius. Thus, introducing the variable $u=vw_0$ in \eqref{FS-L} and by using the formulae for the modified Bessel function for small values of the argument, to the leading order, we obtain
\begin{eqnarray}
	\langle|\varphi(x)|^2\rangle_{b}^{(q,\alpha_0)}&\approx&-\frac{2^{1-2\nu-D/2}}{\pi^{D/2}\Gamma^2(\nu+1)a^{D-1}}\left(\frac{w}{w_0}\right)^{D+2\nu+2}\int_{0}^{\infty}duu^{D+2\nu-1}\frac{\bar{K}_{\nu}(u)}{\bar{I}_{\nu}(u)}\Biggl[\sideset{}{'}\sum_{k=0}^{[q/2]}\cos(2\pi k\alpha_0)\nonumber\\
	&\times&f_{\frac{D}{2}-1}(2rus_k/w_0)
	-\frac{q}{2\pi }\int_{0}^{\infty}dy\frac{h(q,	\alpha_0,y)}{\cosh{(qy)}-\cos{(q\pi)}}\nonumber\\
	&\times&f_{\frac{D}{2}-1}(2ru\cosh(y/2)/w_0)\Biggl] \ .
	\label{FS-asymp-L}
\end{eqnarray}

In the Minkowskian limit, we follow the same procedure as in the R-region. The corresponding result reads,
\begin{eqnarray}
	\langle|\varphi(x)|^2\rangle_{b}^{(q,\alpha_0),(\rm{M})}&=&-\frac{1}{(2\pi)^{D/2}}\int_{m}^{\infty}du(u^2-m^2)^{\frac{D}{2}-1}\Biggl[\sideset{}{'}\sum_{k=0}^{[q/2]}\cos(2\pi k\alpha_0)\nonumber\\
	&\times&f_{\frac{D}{2}-1}(2rs_k\sqrt{u^2-m^2})
	-\frac{q}{2\pi }\int_{0}^{\infty}dy\frac{h(q,	\alpha_0,y)}{\cosh{(qy)}-\cos{(q\pi)}}\nonumber\\
	&\times&f_{\frac{D}{2}-1}(2r\cosh(y/2)\sqrt{u^2-m^2})\Biggl]\frac{1+\beta u}{1-\beta u}e^{-2u(y_0-y)} \ .
	\label{FS-L-M_case}
\end{eqnarray}
It worths to note that the expression above is similar to that of the R-region with $y-y_0$ replaced by $y_0-y$. This similarity is expected since in the Minkowskian limit the VEV is symmetric to the brane.

\begin{figure}[!htb]
	\begin{center}
		\centering
		\includegraphics[scale=0.3]{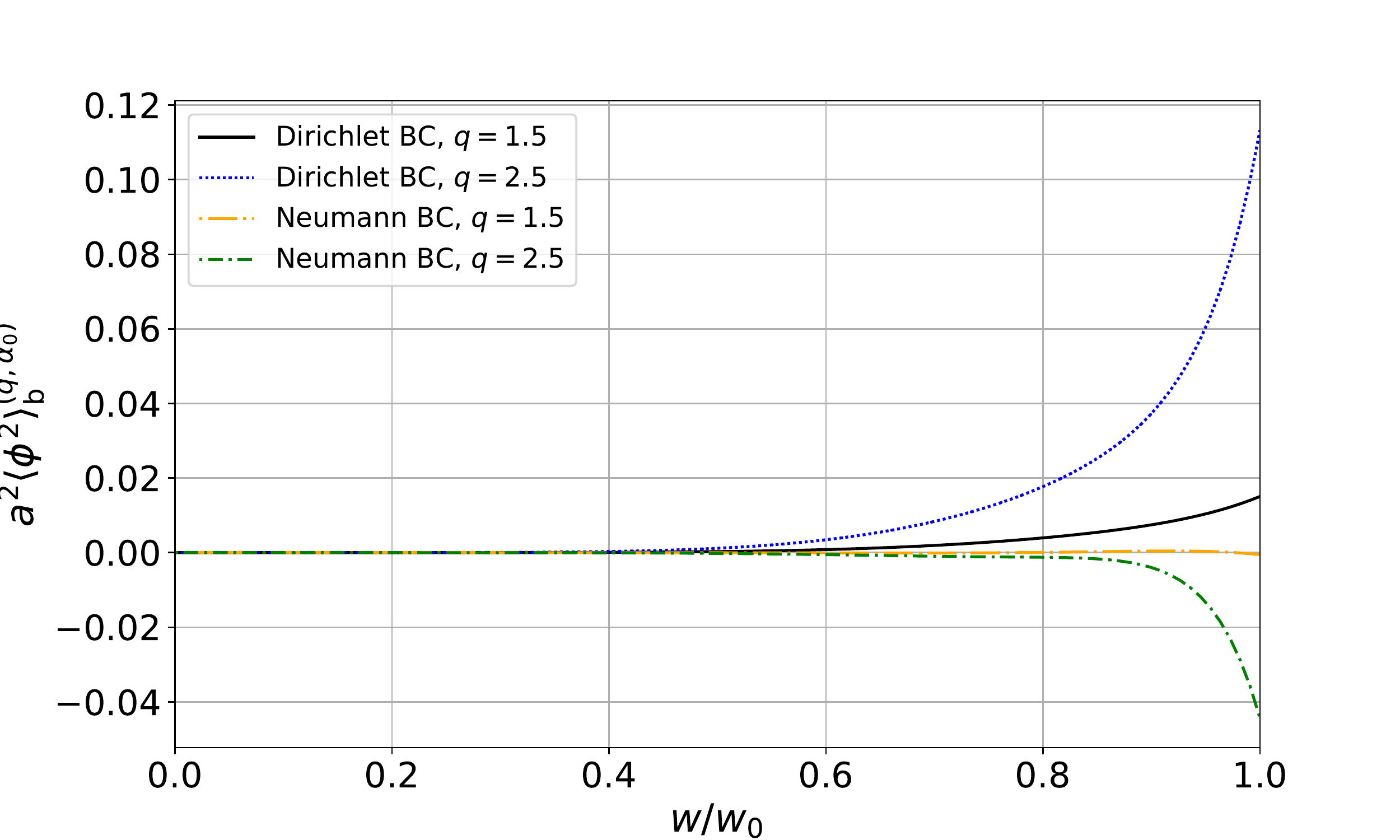}
		\quad
		\includegraphics[scale=0.3]{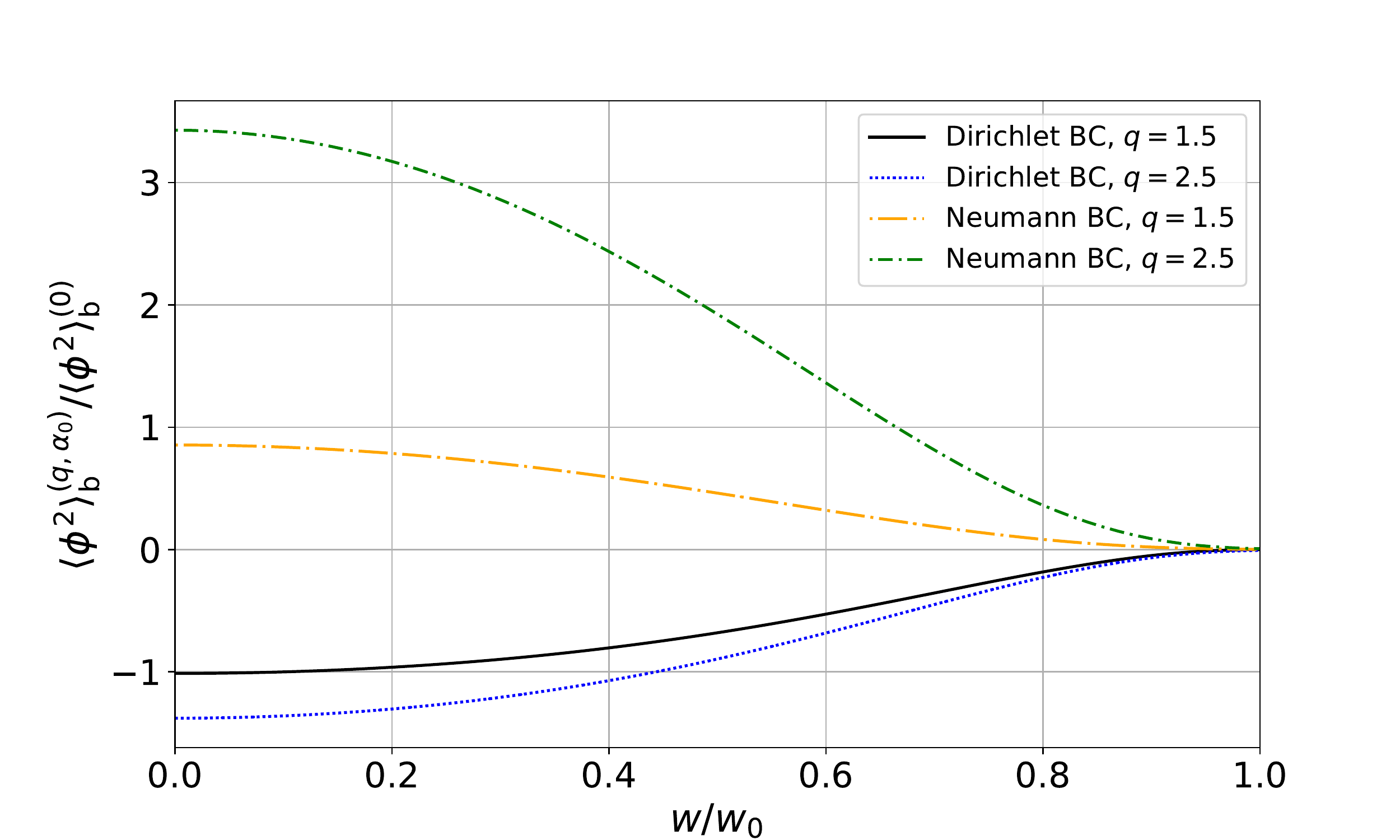}
		\caption{The same as in Fig.\ref{fig1} for the R-region.}
		\label{fig2}
	\end{center}
\end{figure}

The left panel in Fig. \ref{fig2} shows the behaviour of the field squared and right panel the ratio $\langle|\varphi(x)|^2\rangle_{b}^{(q,\alpha_0)}/	\langle|\varphi(x)|^2\rangle_{b}^{(0)}$ as functions of $w/w_0$ for Neumann and Dirichlet boundary conditions and distinct values of $q$. In both plots we have fixed $r/w_0=0.5$ and $\alpha_0=0.4$. From the left panel we can see that the field squared in this region goes zero on the AdS boundary, which is in accordance with the asymptotic formula \eqref{FS-asymp-L}, and it is finite on the brane. Moreover, the right panel shows us that the string-induced contribution is dominant near the AdS boundary for the curves with $q=2.5$, while the pure brane-induced term is more relevant close to the brane for any value of $q$. Note that similar to the L-region, this VEV increases with the string parameter, $q$, but differently from the R-region it is higher for Dirichlet BC.
\\
%

\section{Energy-Momentum Tensor}\label{sec5}
Having obtained the Wightman function and the mean field square, we can proceed to the calculation of the vacuum expectation value of the energy–momentum tensor by making use of the formula developed in \cite{Wagner_20}:
\begin{equation}
	\langle T_{\mu\nu}\rangle=\lim_{x^\prime\rightarrow x}(D_{\mu}D_{\nu'}^{\dagger}+D_{\mu'}^{\dagger}D_{\nu})W(x,x')-2[\xi R_{\mu\nu}+\xi\nabla_{\mu}\nabla_{\nu}-(\xi-1/4)g_{\mu\nu}\nabla_{\alpha}\nabla^{\alpha}]]\langle|\varphi|^2\rangle,
	\label{EM-Tensor-formula}
\end{equation}
where $R_{\mu\nu}=-Dg_{\mu\nu}/a^2$ is the Ricci tensor for the AdS space-time and $D_{\mu}=\nabla_{\mu}+ieA_{\mu}$.\footnote{As it was explained in \cite{Wagner_20}, the second part on the right-hand side of the energy-momentum tensor comes from two distinct contributions: the one proportional to the non-minimum coupling, $\xi$, is a consequence of the variation of the Ricci scalar with respect to the metric tensor; and the second one is obtained through some algebraic manipulations and use of the equation of motion associated with the scalar field.} Similarly to the VEV of the field squared, the VEV of the energy–momentum tensor can be decomposed as
\begin{equation}
	\langle T_{\mu\nu}\rangle_{\rm{b}}=\langle T_{\mu\nu}\rangle_{\rm{b}}^{(0)}+\langle T_{\mu\nu}\rangle_{\rm{b}}^{(q,\alpha_0)} \ .
\end{equation}
As already stressed in beginning of the previous section, here the first term in the right-hand side is also purely induced by the brane, already analyzed in \cite{Saharian:2003qs}. Thus, the analysis below concerns only the second term, which is a new contribution induced by the string and the magnetic flux along its core in the AdS background with a flat boundary.
\subsection{R-region}
Let us start with the R-region, walking through the most important steps of the calculation. The covariant d’Alembertian acting on the squared field in the R-region, Eq. \eqref{FS-R}, gives
\begin{eqnarray}
	\Box\langle|\varphi|^2\rangle_{b}&=&\frac{4w^{D+2}}{(2\pi)^{D/2}a^{D+1}}\int_{0}^{\infty}dpp^{D+1}\frac{\bar{I}_{\nu}(pw_0)}{\bar{K}_{\nu}(pw_0)}\Biggl[\sideset{}{'}\sum_{k=0}^{[q/2]}\cos(2\pi k\alpha_0)S[s_k,f_{\frac{D}{2}}(2rps_k),K_{\nu}(pw)]
	\nonumber\\
	&-&\frac{q}{2\pi }\int_{0}^{\infty}dy\frac{h(q,	\alpha_0,y)}{\cosh{(qy)}-\cos{(q\pi)}}S[\cosh(y/2),f_{\frac{D}{2}}(2rp\cosh(y/2)),K_{\nu}(pw)]\Biggl] \ ,
	\label{EM_Tensor-R}
\end{eqnarray}
where we have introduced the function
\begin{eqnarray}
	S[\gamma,f_{\mu}(x),g(y)]&=&2
	\gamma^2\left[x^2f_{\mu+1}(x)-2f_{\mu}(x)\right]g^2(y)
	\nonumber\\
	&+&f_{\mu-1}(x)\left[(g^\prime(y))^2+\frac{D}{y}g^{\prime
	}(y)g(y)+\left(1+\frac{\nu^2}{y^2}\right)g^2(y)\right] \ ,
\end{eqnarray}
with $\gamma=s_k,\cosh(y/2)$. The function $f_{\mu}(x)$ is already defined in \eqref{func-f}.

For the geometry under consideration, only the $\nabla_{r}\nabla_{w}$ and $\nabla_{\mu}\nabla_{\mu}$ differential operators contribute when acting on the VEV of the field squared.
The remaining contributions come from the electromagnetic covariant derivatives acting on the Wightman function. As to the azimuthal term, it is more convenient to act the $D_{\phi}D_{\phi'}^{\dagger}$ operator in \eqref{W-function_b} for the R-region and \eqref{W-function_full-L} for the L-region, and subsequently take the coincidence limit in all coordinates, including the angular one. Following this procedure, we obtain:
\begin{equation}
	\mathcal{I}(q,\alpha,\chi)=\sum_{n=-\infty}^{\infty}q^2(n+\alpha)^2I_{q|n+\alpha|}(\chi) \ ,
\end{equation}
where $\chi=rr'/2s^2$. This sum can be developed by using the differential equation obeyed by the modified Bessel function. Then we get,
\begin{equation}
	\mathcal{I}(q,\alpha,\chi)=\bigg(\chi^2\frac{d^2}{d\chi^2}+\chi\frac{d}{d\chi}-\chi^2\bigg)\sum_{n=-\infty}^{\infty}I_{q|n+\alpha|}(\chi) \ ,
\end{equation}
where this last sum is given by \cite{Braganca:2014qma}
\begin{equation}
	\sum_{n=-\infty}^{\infty}I_{q|n+\alpha|}(\chi)=\frac{2}{q}\sideset{}{'}\sum_{k=0}^{[q/2]}\cos(2\pi k\alpha_0)e^{\chi\cos(2\pi k/q)}-\frac{1}{\pi}\int_{0}^{\infty}dy\frac{e^{-\chi\cosh(y)}f(q,\alpha_0,y)}{\cosh(qy)-\cos(q\pi)} \ .
\end{equation}

The brane induced contribution in the VEV of the energy-momentum tensor is calculated by making use of the corresponding parts in the Wightman function and VEV of the field squared. After long but straightforward steps, we get (no summation over $\mu$)
\begin{eqnarray}
	\langle T_{\mu}^{\mu}\rangle_{\rm{b}}^{(q,\alpha_0)}&=&-\frac{2w^{D+2}}{(2\pi)^{\frac{D}{2}}a^{D+1}}\int_{0}^{\infty}dpp^{D+1}\frac{\bar{I}_{\nu}(pw_0)}{\bar{K}_{\nu}(pw_0)}\Bigg[\sideset{}{'}\sum_{k=1}^{[q/2]}\cos(2\pi k\alpha_0)G_{\mu}^{\mu}[s_k,f_{\frac{D}{2}}(2rps_k),K_{\nu}(pw)]\nonumber\\
	&-&\frac{q}{2\pi}\int_{0}^{\infty}dy\frac{h(q,\alpha_0,y)}{\cosh(qy)-\cos(q\pi)}G_{\mu}^{\mu}[\cosh(y/2),f_{\frac{D}{2}}(2rp\cosh(y/2)),K_{\nu}(pw)]\Bigg]  \  ,
	\label{EM-Brane}
\end{eqnarray}
with the functions
\begin{eqnarray}
	G_{0}^{0}[\gamma,f_{\mu}(x),g(y)]&=&-2f_{\mu}(x)g^2(y)-\xi_1\Biggl\{2\gamma^2\left[x^2f_{\mu+1}(x)-2f_{\mu}(x)\right]g^2(y)
	\nonumber\\
	&+&f_{\mu-1}(x)\left[(g^\prime(y))^2+\frac{D+4\xi/\xi_1}{y}g^{\prime
	}(y)g(y)+\left(1+\frac{\nu^2}{y^2}\right)g^2(y)\right]\Biggr\} \ , \nonumber\\
	G_{1}^{1}[\gamma,f_{\mu}(x),g(y)]&=&2(4\xi\gamma^2-1)f_{\mu}(x)g^{2}(y)-\xi_1f_{\mu-1}(x)\Biggl[(g^\prime(y))^2+\frac{D+4\xi/\xi_1}{y}g^{\prime
	}(y)g(y)\nonumber\\
	&+&\left(1+\frac{\nu^2}{y^2}\right)g^2(y)\Biggr] \ ,
	\nonumber\\
	G_{2}^{2}[\gamma,f_{\mu}(x),g(y)]&=&-2(4\xi\gamma^2-1)[x^2f_{\mu+1}(x)-f_{\mu}(x)]g^2(y)\nonumber\\
	&-&\xi_1f_{\mu-1}(x)\Biggl[(g^\prime(y))^2+\frac{D+4\xi/\xi_1}{y}g^{\prime
	}(y)g(y)+\left(1+\frac{\nu^2}{y^2}\right)g^2(y)\Biggr] \ ,\nonumber\\
	G_{3}^{3}[\gamma,f_{\mu}(x),g(y)]&=&-2\xi_1\gamma^2\left[x^2f_{\mu+1}(x)-2f_{\mu}(x)\right]g^2(y)
	\nonumber\\
	&+&f_{\mu-1}(x)\Biggl[-(g^\prime(y))^2+\frac{\xi_1D}{y}g^{\prime
	}(y)g(y)+\left(1+\frac{2m^2a^2-\nu^2}{y^2}\right)g^2(y)\Biggr] \ ,
	\label{G-functions}
\end{eqnarray}
where $\xi_1=4\xi-1$. For the components $\mu=4,...,D$, associated to the extra dimensions, we have (no summation over $\mu$) $\langle T_{\mu}^{\mu}\rangle_{\rm{b}}=\langle T_{0}^{0}\rangle_{\rm{b}}$, as a consequence of the symmetry of the problem. Additionally, we have an off-diagonal component\footnote{Note that although the spacetime line element given by \eqref{HDCS} is diagonal, we have the appearance of an off-diagonal component in the scalar field energy-momentum tensor, which is a common feature of similar setups in anti-de Sitter and de Sitter spacetimes with a cosmic string \cite{Wagner_20,BezerradeMello:2009ng}.}:
\begin{equation}
		G_{3}^{1}[\gamma,f_{\mu}(x),g(y)]=-\gamma xf_{\mu}(x)\left[\frac{D\xi_1+4\xi}{y}g^2(y)+2\xi_1g(y)g^\prime(y)\right] \ .
		\label{Off-diagonal}
\end{equation}

Let us now study some limiting cases for the energy density component, $\langle T_{0}^{0}\rangle_{\rm{b}}^{(q,\alpha_0)}$. For a massless conformal quantum field, we have $\nu=1/2$, and using the corresponding modified Bessel functions, the energy density is given by
\begin{eqnarray}
	\langle T_{0}^{0}\rangle_{\rm{b}}^{(q,\alpha_0)}&=&-\frac{4w^{D+1}}{(2\pi)^{\frac{D}{2}}a^{D+1}D}\int_{0}^{\infty}dpp^{D}e^{-(2w-w_0)p}\frac{(2A_0-B_0) \sinh (pw_0)+2 B_0 pw_0 \cosh (pw_0)}{2A_0-B_0 (1+2pw_0)} \nonumber\\
	&\times&\Biggl\{\sideset{}{'}\sum_{k=1}^{[q/2]}\cos(2\pi k\alpha_0)\Biggl[s_k^2\Big((2rps_k)^2f_{\frac{D}{2}+1}(2rps_k)-2f_{\frac{D}{2}}(2rps_k)\Big)-Df_{\frac{D}{2}}(2rps_k)\nonumber\\
	&+&f_{\frac{D}{2}-1}(2rps_k)\Biggr]-\frac{q}{2\pi}\int_{0}^{\infty}dy\frac{h(q,\alpha_0,y)}{\cosh(qy)-\cos(q\pi)}\Biggl[\cosh^2(y/2)\Big((2rp\cosh(y/2))^2\nonumber\\
	&\times&f_{\frac{D}{2}+1}(2rp\cosh(y/2))-2f_{\frac{D}{2}}(2rp\cosh(y/2))\Big)	-Df_{\frac{D}{2}}(2rp\cosh(y/2))\nonumber\\
	&+&
	f_{\frac{D}{2}-1}(2rp\cosh(y/2))\Biggr]\Biggr\} \  .
	\label{EM-MSF-R}
\end{eqnarray}
Note that in contrast to the vanishing pure brane-induced energy density in the R-region analyzed in \cite{Saharian:2003qs}, the corresponding contribution resulting from the presence of the cosmic string and its magnetic flux along its core given by the expression above is non-zero.

For distant points from the brane, $w/w_0\gg1$, we introduce a new variable $u=pw$ in \eqref{EM-Brane} for the energy density component and by making use of the formulae for the modified Bessel functions for small values of the argument \cite{Abra}, with the assumption that $A_0-\nu B_0\neq0$, to the leading order, we get
\begin{eqnarray}
	\langle T_{0}^{0}\rangle_{\rm{b}}^{(q,\alpha_0)}&\approx&-\frac{2^{2-2\nu-D/2}}{\pi^{\frac{D}{2}}\Gamma(\nu)\Gamma(\nu+1)a^{D+1}}\left(\frac{w_0}{w}\right)^{2\nu}\int_{0}^{\infty}dpp^{2\nu+D+1}\left(\frac{A_0+\nu B_0}{A_0-\nu B_0}\right)\nonumber\\
	&\times&
	\Bigg[\sideset{}{'}\sum_{k=1}^{[q/2]}\cos(2\pi k\alpha_0)G_{0}^{0}[s_k,f_{\frac{D}{2}}(2rps_k/w),K_{\nu}(p)]-\frac{q}{2\pi}\int_{0}^{\infty}dy\frac{h(q,\alpha_0,y)}{\cosh(qy)-\cos(q\pi)}\nonumber\\
	&\times&G_{0}^{0}[\cosh(y/2),f_{\frac{D}{2}}(2rp\cosh(y/2)/w),K_{\nu}(p)]\Bigg]  \  .
	\label{EM-asymp-R}
\end{eqnarray}
In the Minkowskian limit, $a\rightarrow\infty$, the energy density reads,
\begin{eqnarray}
	\langle T_{0}^{0}\rangle_{\rm{b}}^{(q,\alpha_0),\rm(M)}&=&\frac{2}{(2\pi)^{\frac{D}{2}}}\int_{m}^{\infty}du(u^2-m^2)^{\frac{D}{2}-1}\Biggl\{\sideset{}{'}\sum_{k=1}^{[q/2]}\cos(2\pi k\alpha_0)\Biggl[u^2\Big(\xi_1f_{\frac{D}{2}-1}(2rs_k\sqrt{u^2-m^2})\nonumber\\
	&+&f_{\frac{D}{2}}(2rs_k\sqrt{u^2-m^2})\Big)-m^2f_{\frac{D}{2}}(2rs_k\sqrt{u^2-m^2})\Biggr]\nonumber\\
	&-&\frac{q}{2\pi}\int_{0}^{\infty}dy\frac{h(q,\alpha_0,y)}{\cosh(qy)-\cos(q\pi)}\Biggl[u^2\Big(\xi_1f_{\frac{D}{2}-1}(2r\cosh(y/2)\sqrt{u^2-m^2})\nonumber\\
	&+&f_{\frac{D}{2}}(2r\cosh(y/2)\sqrt{u^2-m^2})\Big)-m^2f_{\frac{D}{2}}(2r\cosh(y/2)\sqrt{u^2-m^2})\Bigg]\Biggr\}\nonumber\\
	&\times&\frac{1+\beta u}{1-\beta u}e^{-2u(y-y_0)} \ .
	\label{E-density-M_case}
\end{eqnarray}

\begin{figure}[!htb]
	\begin{center}
		\centering
		\includegraphics[scale=0.3]{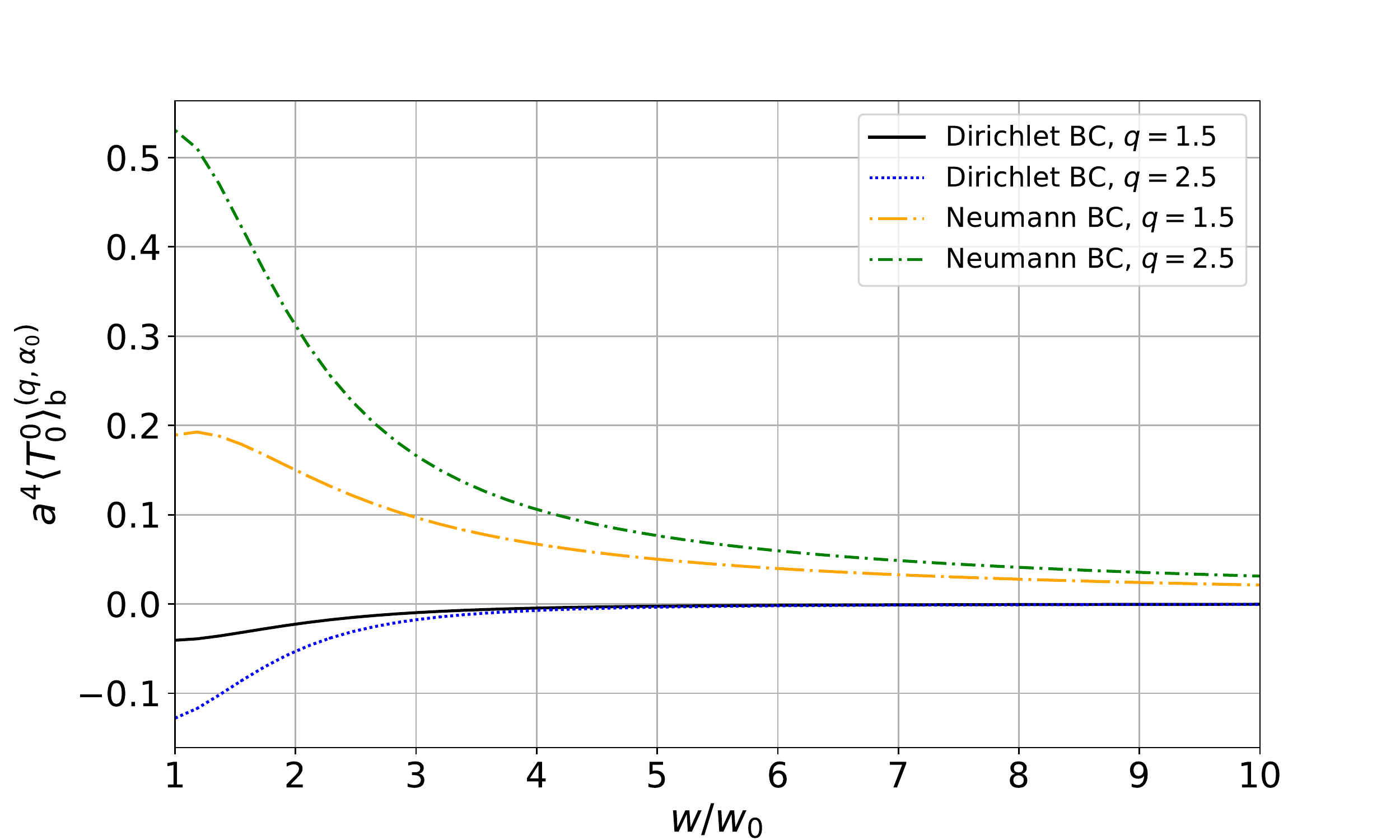}
		\quad
		\includegraphics[scale=0.3]{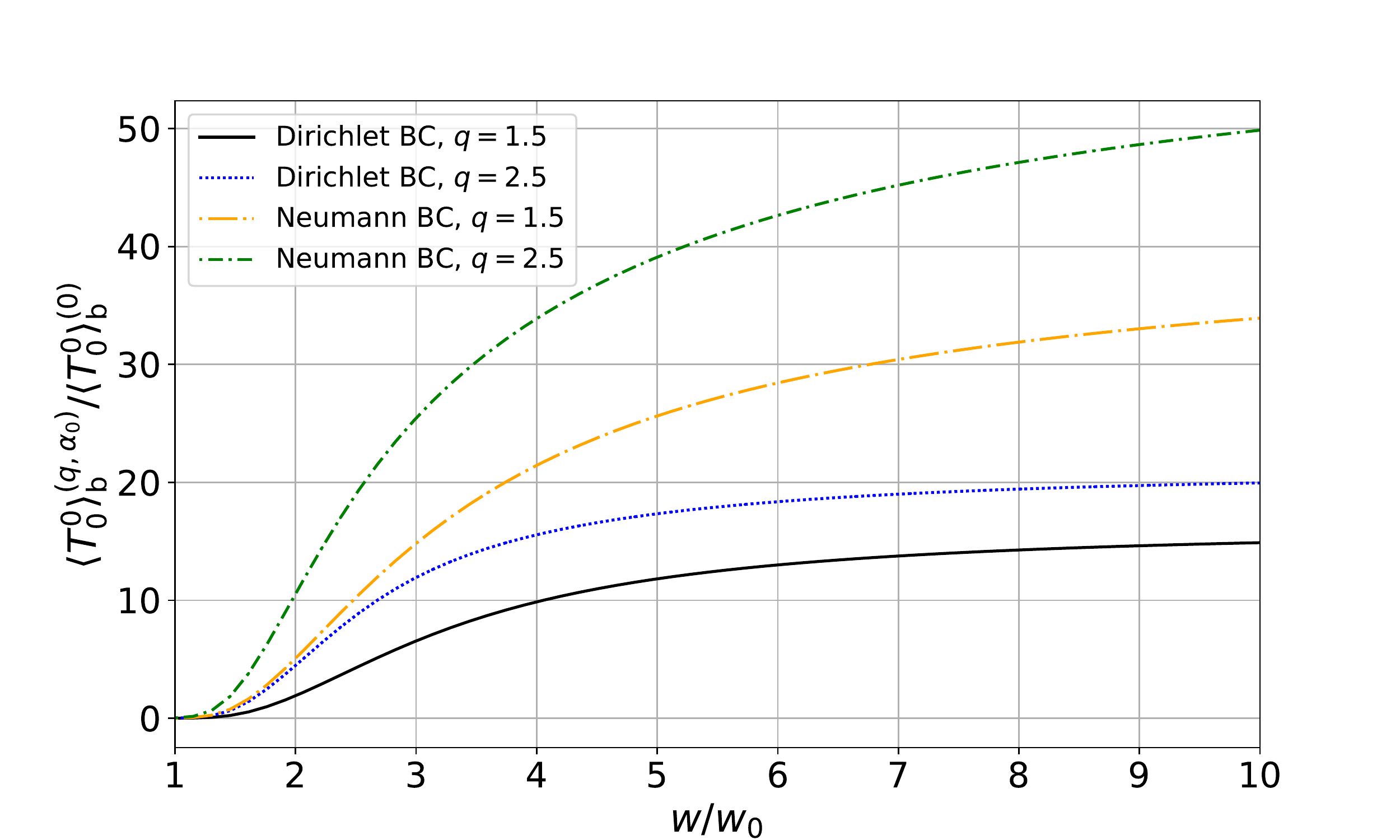}
		\caption{The behaviour of the energy density as  function of $w/w_0$ is shown for different values of $q$ and for Dirichlet and Neumann boundary conditions (left panel). The right panel exhibits the ratio between the contribution $\langle T_{0}^{0}\rangle_{\rm{b}}^{(q,\alpha_0)}$ and $\langle T_{0}^{0}\rangle_{\rm{b}}^{(0)}$ also as a function of $w/w_0$ for distinct values $q$ and Dirichlet and Neumann boundary conditions. The graphs are plotted for a minimally coupled massless scalar field with fixed parameters $r/w_0=0.5$ and $\alpha_0=0.4$.}
		\label{fig3}
	\end{center}
\end{figure}

In Fig. \ref{fig3} the energy density induced by the string and its magnetic flux, $\langle T_{0}^{0}\rangle_{\rm{b}}^{(q,\alpha_0)}$ (left panel) and the ratio $\langle T_{0}^{0}\rangle_{\rm{b}}^{(q,\alpha_0)}/\langle T_{0}^{0}\rangle_{\rm{b}}^{(0)}$ (right panel) are displayed as functions of $w/w_0$. Both graphs are plotted for a minimally coupled massless scalar field in $D=3$ with fixed parameters $r/w_0=0.5$ and $\alpha_0=0.4$.  The curves exhibited correspond to different values of $q$, considering Dirichlet ($\beta=0$) and Neumann ($\beta\rightarrow\infty$) boundary conditions. The left panel shows us that the intensity of the energy density increases with the parameter $q$ associated with the string's angle deficit. Moreover, the energy density goes to zero for large distances from the brane, and according to our asymptotic analysis behaves as $(w_0/w)^{2\nu}$, and it is finite on the brane. We can also observe that in the region under consideration the energy density is positive for Neumann BC and negative for Dirichlet BC and that the intensities of the VEVs are higher for Neumann BC by comparing curves with same $q$. On the other hand, from the right panel we can read that the contribution induced by the string and its magnetic flux, $\langle T_{0}^{0}\rangle_{\rm{b}}^{(q,\alpha_0)}$, is negligible compared with the pure brane-induced one, $\langle T_{0}^{0}\rangle_{\rm{b}}^{(0)}$, for points close to the brane. On the other hand, for distant points from the brane the former dominates in the total VEV, $\langle T_{0}^{0}\rangle_{\rm{b}}$.
\\
\subsection{L-region}
Following the same procedure as in the R-region, the energy-momentum tensor in the L-region reads,
\begin{eqnarray}
	\langle T_{\mu}^{\mu}\rangle_{\rm{b}}^{(q,\alpha_0)}&=&-\frac{2w^{D+2}}{(2\pi)^{\frac{D}{2}}a^{D+1}}\int_{0}^{\infty}dpp^{D+1}\frac{\bar{K}_{\nu}(pw_0)}{\bar{I}_{\nu}(pw_0)}\Bigg[\sideset{}{'}\sum_{k=1}^{[q/2]}\cos(2\pi k\alpha_0)G_{\mu}^{\mu}[s_k,f_{\frac{D}{2}}(2rps_k),I_{\nu}(pw)]\nonumber\\
	&-&\frac{q}{2\pi}\int_{0}^{\infty}dy\frac{h(q,\alpha_0,y)}{\cosh(qy)-\cos(q\pi)}G_{\mu}^{\mu}[\cosh(y/2),f_{\frac{D}{2}}(2rp\cosh(y/2)),I_{\nu}(pw)]\Bigg]  \  ,
	\label{EM_brane-L}
\end{eqnarray}
with the functions $G_{\mu}^{\mu}$ already defined in \eqref{G-functions}.
For the components $\mu=4,...,D$, associated to the extra dimensions, in the L-region we also have (no summation over $\mu$) $\langle T_{\mu}^{\mu}\rangle_{\rm{b}}=\langle T_{0}^{0}\rangle_{\rm{b}}$ and the off-diagonal component reads the same as in \eqref{Off-diagonal}. Moreover, note that the energy-momentum tensor in the L-region is obtained from  \eqref{EM-Brane} with the replacements $I\rightarrow K$ and
$K\rightarrow I$ of the modified Bessel functions.

For a massless conformal quantum field, we have $\nu=1/2$, and using the corresponding modified Bessel functions, the energy density is given by
\begin{eqnarray}
	\langle T_{0}^{0}\rangle_{\rm{b}}^{(q,\alpha_0)}&=&-\frac{2w^{D+1}}{(2\pi)^{\frac{D}{2}}a^{D+1}D}\int_{0}^{\infty}dpp^{D}e^{-pw_0}\frac{ 2A_0-B_0(1+2pw_0)}{(2 A_0-B_0) \sinh (pw_0)+2 B_0 pw_0 \cosh (pw_0)} \nonumber\\
	&\times&\Biggl\{\sideset{}{'}\sum_{k=1}^{[q/2]}\cos(2\pi k\alpha_0)\Biggl[\left[2s_k^2\left((2rps_k)^2f_{\frac{D}{2}+1}(2rps_k)-2f_{\frac{D}{2}}(2rps_k)\right)-2Df_{\frac{D}{2}}(2rps_k)\right]\nonumber\\
	&\times&\sinh^2(pw)+f_{\frac{D}{2}-1}(2rps_k)[2\cosh^2(pw)-1]\Biggr]-\frac{q}{2\pi}\int_{0}^{\infty}dy\frac{h(q,\alpha_0,y)}{\cosh(qy)-\cos(q\pi)}\nonumber\\
	&\times&\Biggl[\Bigl[2\cosh^2(y/2)\left((2rp\cosh(y/2))^2f_{\frac{D}{2}+1}(2rp\cosh(y/2))-2f_{\frac{D}{2}}(2rps_k)\right)\nonumber\\
	&-&2Df_{\frac{D}{2}}(2rp\cosh(y/2))\Bigr]\sinh^2(pw)+f_{\frac{D}{2}-1}(2rp\cosh(y/2))[2\cosh^2(pw)-1]\Biggr]\Biggr\} .
	\label{EM-MSF-L}
\end{eqnarray}

For points near the AdS boundary, $w\ll w_0$, we introduce the variable $u=vw_0$ in \eqref{EM_brane-L} for the energy density component and by using the corresponding expressions for the modified Bessel functions for small values of the argument, to the leading order, we get
\begin{eqnarray}
	\langle T_{0}^{0}\rangle_{\rm{b}}^{(q,\alpha_0)}&\approx&-\frac{2^{1-2\nu-D/2}}{\pi^{\frac{D}{2}}a^{D+1}}\left(\frac{w}{w_0}\right)^{D+2\nu}\frac{D+2\nu-4\xi(D+2\nu+1)}{\Gamma(\nu)\Gamma(\nu+1)}\int_{0}^{\infty}dpp^{D+2\nu-1}\frac{\bar{K}_{\nu}(pw_0)}{\bar{I}_{\nu}(pw_0)}\nonumber\\
	&\times&\Bigg[\sideset{}{'}\sum_{k=1}^{[q/2]}\cos(2\pi k\alpha_0)f_{\frac{D}{2}-1}(2rps_k)-\frac{q}{2\pi}\int_{0}^{\infty}dy\frac{h(q,\alpha_0,y)}{\cosh(qy)-\cos(q\pi)}\nonumber\\
	&\times&f_{\frac{D}{2}-1}(2rp\cosh(y/2))\Bigg]  \  .
	\label{EM-asymp-L}
\end{eqnarray}

In the Minkowskian limit, $a\rightarrow\infty$, the energy density reads,
\begin{eqnarray}
	\langle T_{0}^{0}\rangle_{\rm{b}}^{(q,\alpha_0),\rm(M)}&=&\frac{2}{(2\pi)^{\frac{D}{2}}}\int_{m}^{\infty}du(u^2-m^2)^{\frac{D}{2}-1}\Biggl\{\sideset{}{'}\sum_{k=1}^{[q/2]}\cos(2\pi k\alpha_0)\Biggl[u^2\Big(\xi_1f_{\frac{D}{2}-1}(2rs_k\sqrt{u^2-m^2})\nonumber\\
	&+&f_{\frac{D}{2}}(2rs_k\sqrt{u^2-m^2})\Big)-m^2f_{\frac{D}{2}}(2rs_k\sqrt{u^2-m^2})\Biggr]\nonumber\\
	&-&\frac{q}{2\pi}\int_{0}^{\infty}dy\frac{h(q,\alpha_0,y)}{\cosh(qy)-\cos(q\pi)}\Biggl[u^2\Big(\xi_1f_{\frac{D}{2}-1}(2r\cosh(y/2)\sqrt{u^2-m^2})\nonumber\\
	&+&f_{\frac{D}{2}}(2r\cosh(y/2)\sqrt{u^2-m^2})\Big)-m^2f_{\frac{D}{2}}(2r\cosh(y/2)\sqrt{u^2-m^2})\Bigg]\Biggr\}\nonumber\\
	&\times&\frac{1+\beta u}{1-\beta u}e^{-2u(y_0-y)} \ .
	\label{E-density-M_case-L}
\end{eqnarray}
It is noteworthy that the energy density, similar to the VEV of the squared field, is also similar to that of the R-region with $y-y_0$ replaced by $y_0-y$, which is also expected since in the Minkowskian limit the VEV is symmetric to the brane.

\begin{figure}[!htb]
	\begin{center}
		\centering
		\includegraphics[scale=0.3]{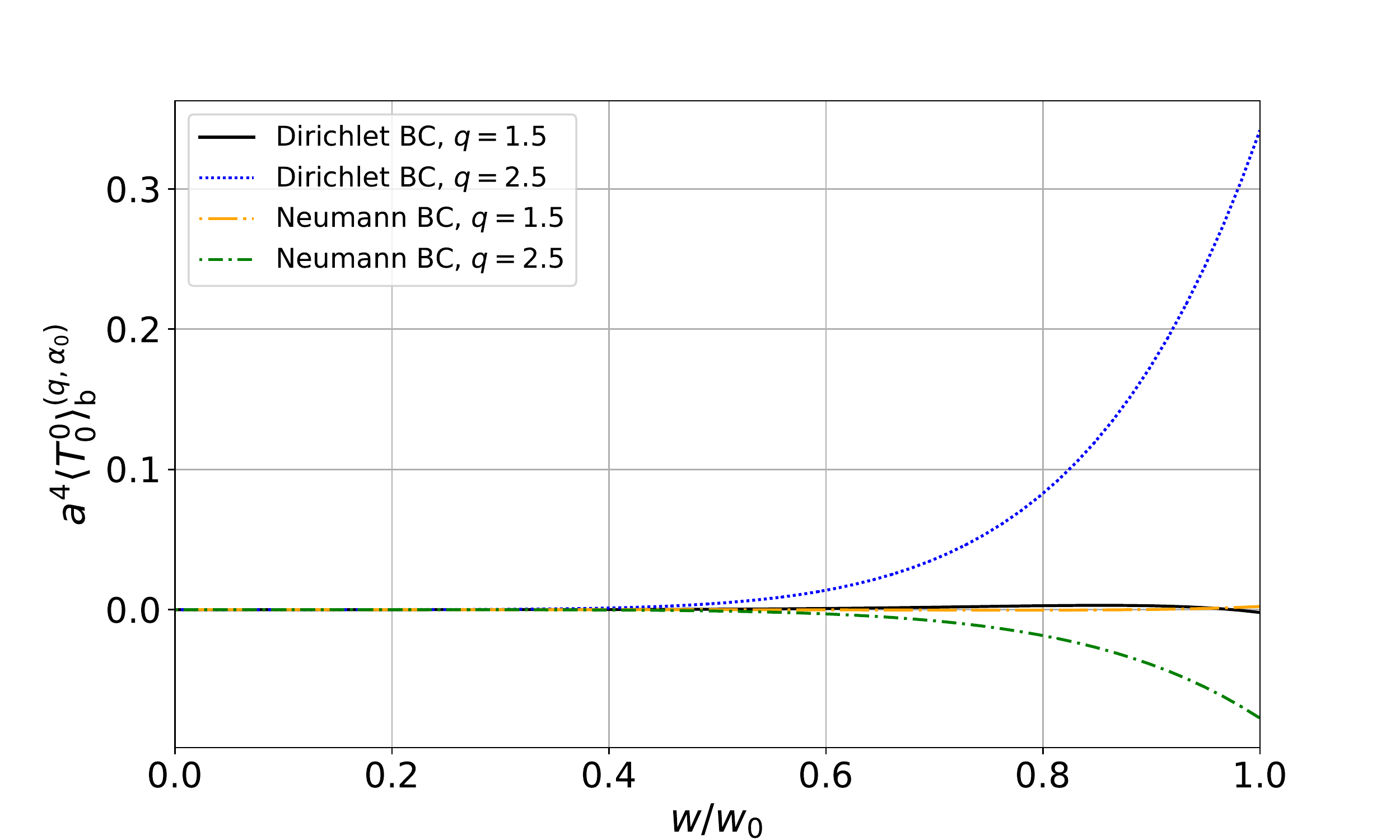}
		\quad
		\includegraphics[scale=0.3]{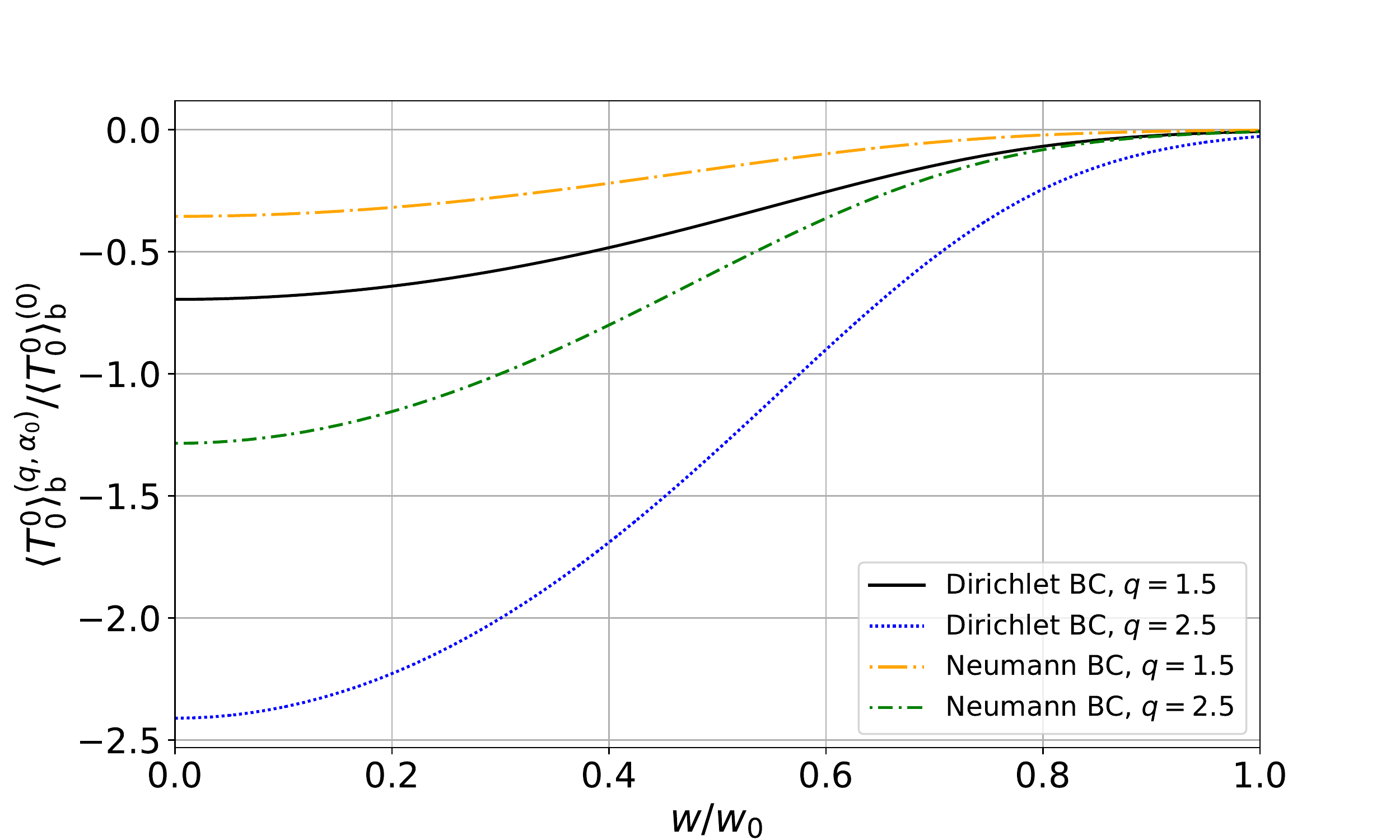}
		\caption{The energy density $\langle T_{0}^{0}\rangle_{\rm{b}}^{(q,\alpha_0)}$ the ratio $\langle T_{0}^{0}\rangle_{\rm{b}}^{(q,\alpha_0)}/\langle T_{0}^{0}\rangle_{\rm{b}}^{(0)}$ are plotted as functions of $w/w_0$ for different values $q$ and Dirichlet and Neumann boundary conditions. The graphs are plotted for a minimally coupled massless scalar field in $D=3$ with fixed parameters $r/w_0=0.5$ and $\alpha_0=0.4$.}
		\label{fig4}
	\end{center}
\end{figure}

In Fig. \ref{fig4} we present two plots showing the energy density induced in the presence of the cosmic string and its magnetic flux (left panel) and the ratio $\langle T_{0}^{0}\rangle_{\rm{b}}^{(q,\alpha_0)}/\langle T_{0}^{0}\rangle_{\rm{b}}^{(0)}$ (right panel) as functions of $w/w_0$ for different values of the string parameter, $q$, considering Dirichlet and Neumann boundary conditions. From both plots we can observe an inversion of the behaviour found in the R-region; the energy density is negative for Neumann BC and positive for Dirichlet BC, being the intensities higher for the latter by comparing the curves with same value of $q$. Moreover, we can see that in this region the energy density goes to zero near the AdS boundary with $(w/w_0)^{D+2\nu}$, according to the corresponding asymptotic expression \eqref{EM-asymp-L}, and it is finite on the brane. From the right panel we can read that the pure brane-induced contribution dominates in the total VEV for point close to the brane. On the other side, for points close to the AdS boundary the contribution induced by the string and its magnetic flux dominates in the total VEV, depending on the deficit angle parameter, $q$.
\\

\section{Application to Randall-Sundrum model}\label{sec6}
The results given in the previous section can be applied to the investigation of the cosmic string induced effects in $Z_2$-symmetric braneworlds models. Specifically, they can be applied to the Randall–Sundrum model with a single brane (RSII) \cite{RSI,RSII}. In the RSII model, the universe is perceived as a $Z_2$-symmetric positive tension brane in 5-dimensional AdS spacetime and the negative cosmological constant in the bulk is the only contribution to the curvature. Nevertheless, the majority of scenarios that are motivated by string theories postulate the existence of additional bulk fields, including scalar fields. For the setup under consideration, the cosmic string is perpendicular to the brane and the corresponding background contains two copies of the R-region that are identified by the $Z_2$-symmetry ($y-y_0\longleftrightarrow y_0-y$) with respect to the brane located at $y=y_0$. The corresponding line element is given by
\begin{equation}
	ds^2=e^{-2|y-y_0|/a}\left[dt^2-dr^2-r^2d\phi^2-dz^2-\sum_{i=4}^{D}(dx^{i})^2\right]-dy^2 \ ,
\end{equation}
where, as before, $-\infty<y<\infty$ and $0\le\phi\le2\pi/q$. It should be noted that for an observer located at $y=y_0$, the line element above is trivially reduced to the standard line element that characterizes a cosmic string in $(D+1)$-dimensional flat spacetime. The boundary conditions on the bulk field are imposed at the location of the brane and obtained by integration of the field equations about $y=y_0$. By following a similar procedure to that used in \cite{Saharian:2003qs,Gherghetta2000,Flachi2001} for a single brane model it can be seen
that for fields even under the reflection with respect to the brane (untwisted scalar field) the boundary condition is of the Robin type \eqref{RBC} with the coefficient
\begin{equation}
	\beta=-\frac{1}{c_b+4D\xi/a} \ ,
	\label{RS-coefficient}
\end{equation}
where $c_b$ is the brane mass term and it comes from the part of the action located on the brane, $S_b=-c_b\int d^Dxdy\sqrt{|g|}\delta(y-y_0)\varphi^2/2$. In the particular case of a minimally coupled field and $c_b=0$, the boundary condition is reduced to the Neumann one ($\beta\rightarrow\infty$). On the other hand, for fields odd under reflection (twisted scalar fields), the boundary condition takes the Dirichlet form. In the $Z_2$-symmetric models the region of integration over the coordinate $y$ ranges from $-\infty$ to $+\infty$ and it results in the appearance of an additional 1/2 factor in the normalization coefficient when compared to the one we have obtained previously for the R-region, which has the interval $0\le y<\infty$ (with $y_0=0$). Therefore, the formulas for the VEVs of the field squared and the energy-momentum tensor induced by a cosmic string in the generalized RSII model are obtained from those expressions presented in the previous sections by putting $w_0=a$ with an additional factor of 1/2.

\section{Conclusions}\label{sec:Conc} 
In this paper we have investigated the vacuum polarization effects induced by a cosmic string carrying a magnetic flux along its core in the background of $(D+1)$-dimensional AdS spacetime with a planar brane parallel to the string core, which divides the background in two regions. In order to obtain the bosonic normal modes in this setup in both regions we have solved the gauge invariant Klein-Gordon equation with a curvature coupling and assumed that on the brane the field operator obeys the Robin boundary condition. Having obtained the positive and negative energy modes in each regions, we have constructed the Wightman function in closed form for R-region \eqref{W-function_b-3} and L-region \eqref{W-function_b-3-L}. By directly taking the coincidence limit in the Wightman function, in the section \ref{sec4} we have calculated the VEV of the field squared induced by the string and its magnetic flux, $\langle|\varphi|^2\rangle_{b}^{(q,\alpha_0)}$, for both regions. In the R-region the VEV of the squared field is given by Eq. \eqref{FS-R}. In Eq. \eqref{FS_MSF} we have presented this quantity for a conformal massless quantum scalar. In the asymptotic limit of large distances from the brane, $w/w_0\gg1$, the string-induced VEV of the field squared  decays as $(w_0/w)^{2\nu}$ and it is given by \eqref{FS-asymp-R}. The Minkowskian limit has been also analyzed and the corresponding expression is givem in Eq. \eqref{FS-R-M_case}. We also have shown that $\langle|\varphi|^2\rangle_{b}^{(q,\alpha_0)}$ is finite on the boundary; moreover, comparing this quantity with the corresponding VEV induced by the brane only,  $\langle|\varphi|^2\rangle_{b}^{0}$, we have noticed by the graph exhibited in the right panel of Fig. \ref{fig1} that near the brane the total VEV of the field squared is dominated by the latter; however, for points more distant  from the brane the string-induced field squared becomes more relevant. In fact the pure brane-induced VEV of the field squared is divergent on the brane \cite{Saharian:2003qs}. In the L-region the VEV of the field squared is given in \eqref{FS-L}. Its conformal massless scalar field case is presented in \eqref{FS-MSF-L}. For points close to the AdS boundary the asymptotic expression is given \eqref{FS-asymp-L} and it is shown that this quantity decreases with $(w/w_0)^{D+2\nu+2}$. The Minkowskian limit of string-induced VEV of the field squared was also studied in this region and it is similar to that one for the R-region with $y-y_0$ replaced by $y_0-y$. In this region, we also analyzed numerically the ratio between the VEV of the field squared induced by the cosmic string and the corresponding quantity induced by the brane, $\langle|\varphi|^2\rangle_{b}^{(q,\alpha_0)} /\langle|\varphi|^2\rangle_{b}^{(0)}$. This result is displayed in Fig. \ref{fig2}. There we observe in the right panel, that the pure brane-induced VEV of the field squared is dominant in the total VEV of the field squared, while for points distant from the brane  the most relevant contribution comes from string-induced part. Also in the L-region, the pure brane-induced VEV of the field squared is divergent on the brane.

In the section \ref{sec5} we have presented our results for the VEV of the energy-momentum tensor, analysing only the cosmic string induced term, $\langle T_{\mu\nu}\rangle_{b}^{(q,\alpha_0)}$, which is a new contribution. The corresponding expression for the R-region is given in \eqref{EM-Brane}. All the diagonal components are nonzero and an off-diagonal component is also present. We have studies the energy density component for some limiting cases. The expression for a conformal massless scalar field case is given in \eqref{EM-MSF-R} is nonzero and it is in clear contrast with the pure brane-induced term, which is zero in this particular case. For distant points of the brane this VEV decreases as $(w_0/w)^{2\nu}$ and is presented in \eqref{EM-asymp-R}. Moreover, the Minkowskian limit is also analyzed and the corresponding expression given by \eqref{E-density-M_case}. In the left panel of Fig. \ref{fig3} we display the behavior of the string-induced energy density, $\langle T_{0}^{0}\rangle_{\rm{b}}^{(q,\alpha_0)}$, as function of $w/w_0$. We show that this quantity is finite on the brane and reinforce its decay for points distant from the brane; moreover, in the right panel, we exhibit the behavior of the ratio $\langle T_{0}^{0}\rangle_{\rm{b}}^{(q,\alpha_0)}/\langle T_{0}^{0}\rangle_{\rm{b}}^{(0)}$. We observe that the brane-induced VEV of the energy density is more intense than the string-induced one near the brane. However the situation changes in the opposite situation. In fact the brane-induced energy density is divergent on the brane  \cite{Saharian:2003qs}.  We also have analyzed the VEV of the energy-momentum tensor in the L-region \eqref{EM_brane-L}, which is obtained from the one for the R-region with the replacements $I\rightarrow K$ and $K\rightarrow I$ of the modified Bessel functions. For a massless scalar field the energy density component is given in \eqref{EM-MSF-L}. For point near the AdS boundary the energy density goes to zero as $(w/w_0)^{D+2\nu}$ as is shown in \eqref{EM-asymp-L}. Furthermore, in the Minkowskian limit the energy density \eqref{E-density-M_case-L} is similar to the corresponding VEV for the R-region with $y-y_0$ replaced by $y_0-y$. In this region, we also have evaluated numerically the behavior of the string-induced VEV of the energy density in the left panel of Fig. \ref{fig4}. We show that it is finite on the brane. On the right panel, we plotted the behavior of the ratio $\langle T_{0}^{0}\rangle_{\rm{b}}^{(q,\alpha_0)}/\langle T_{0}^{0}\rangle_{\rm{b}}^{(0)}$. Again, we can notice that near the brane the total energy density is dominated by the brane-induced part; however for points far from the brane, the string-induced contribution is dominant.
Finally, in the section \ref{sec6} we have applied the results found for the R-region to study the cosmic string induced effects in the generalized Randall-Sundrum model with a single brane. By integrating the field equations about the brane location, $y_0$, the boundary conditions in this $Z_2$-symmetric model is of the Robyn type with coefficient given by \eqref{RS-coefficient} for a field even under reflection with respect to the brane and it is reduced to the Neumann boundary condition in the case of a minimally coupled field, $\xi=0$, and zero brane mass term, $c_b=0$. On the other hand for a field odd we get the Dirichlet boundary condition. The VEVs of the field squared and the energy-momentum tensor induced by a cosmic string and its magnetic flux in the RSII model are then obtained from those found in the section \ref{sec5} by directly putting $w_0=a$ with an additional factor 1/2.

\section*{Acknowledgment}
The authors are grateful to H. F. Mota and A. A. Saharian for helpful discussions during the development of this work. W.O.S is supported under grant 2022/2008, Paraíba State Research Foundation (FAPESQ). E.R.B.M is partially supported by CNPq under Grant no 301.783/2019-3.

\end{document}